\documentclass[twocolumn,floatfix,superscriptaddress,aps,prl]{revtex4-2}
\pdfoutput=1
\usepackage[utf8]{inputenc}
\usepackage[english]{babel}
\usepackage[T1]{fontenc}
\usepackage{amsmath}
\usepackage{hyperref}
\usepackage[normalem]{ulem}

\usepackage{graphicx}
\usepackage{amsmath,mathtools}
\usepackage{amssymb}
\usepackage{bm}
\usepackage{color}
\usepackage{bbold}
\usepackage{mathtools}

\DeclarePairedDelimiterX\braket[2]{\langle}{\rangle}{#1\,\delimsize\vert\,\mathopen{}#2}

\usepackage{tikz}
\usepackage{quantikz}
\usepackage{lipsum}

\begin{document}

\title{Upper bound on \texorpdfstring{$T_c$\ }\ in a strongly coupled electron-boson superconductor}

\author{Nikolay V. Gnezdilov}
\email{Nikolay.Gnezdilov@dartmouth.edu}

\author{Rufus Boyack}
\email{Rufus.Boyack@dartmouth.edu}
\affiliation{Department of Physics and Astronomy, Dartmouth College, Hanover, New Hampshire 03755, USA}

\begin{abstract}
Migdal-Eliashberg theory of boson-mediated superconductivity contains a $\sqrt{\lambda}$ divergence in the critical temperature $T_c$ for strong electron-boson coupling $\lambda$. In the conventional Migdal-Eliashberg theory, the strong-coupling regime can be accessed only in the limit that $\lambda_E = \lambda \, \omega_D/\varepsilon_F\ll1$, where $\omega_D$ is the Debye frequency and $\varepsilon_F$ is the Fermi energy. Here we go beyond this restriction in the context of the two-dimensional Yukawa-SYK (Y-SYK) model,  which is solvable for arbitrary values of $\lambda_E$. We find that $T_c\approx 0.18 \,\omega_D \sqrt{\lambda}$ for large $\lambda$, provided $\lambda_E$ remains small, and crosses over to a universal value of $T_c \approx 0.04\, \varepsilon_F$  for large $\lambda_E$. The saturation of $T_c$ is due to a self-consistent account of the boson dynamics for large $\lambda_E$ and remains valid provided the vertex corrections are negligible. Depending on the value of $\lambda$, this self-consistent approach leads to pairing that describes multiple classes of quantum critical electronic systems. These results demonstrate how the $\sqrt{\lambda}$ growth of $T_c$ in Migdal-Eliashberg theory saturates to a universal value independent of $\lambda$ and $\omega_D$, providing an upper bound on the critical temperature at strong electron-boson coupling. 
\end{abstract}

\maketitle

{\it Introduction.---} 
The Migdal-Eliashberg theory~\cite{Migdal1958, Eliashberg1960, Eliashberg1961} of boson-mediated superconductivity is extremely successful in describing the properties of many superconducting materials~\cite{Parks1, Carbotte1990, Bennemann}. In conventional metals, the bosonic ``glue'' for Cooper-pair formation is provided by the lattice vibrations -- phonons. Migdal-Eliashberg theory explicitly includes boson dynamics, which makes the theory more versatile and more broadly applicable than the standard Bardeen-Cooper-Schrieffer approach~\cite{Bardeen1957a, Bardeen1957, Bardeen1973}. In Migdal-Eliashberg theory, bosons are typically assumed to be much slower than electrons~\cite{Eliashberg1960}, which means $\omega_D/\varepsilon_F \ll 1$ where $\omega_D$ is the Debye frequency and $\varepsilon_F$ is the Fermi energy. This adiabatic approximation has several implications. First, it allows us to consider large electron-boson coupling $\lambda$ if the parameter $\lambda_E = \lambda \omega_D/\varepsilon_F$ is small: in the limit of $\lambda \to \infty$ with $\omega_D/\varepsilon_F\to 0$, there is a $\sqrt{\lambda}$ divergence of the critical temperature $T_c$~\cite{Allen1975Transition, Allen1983Theory, Kresin1984Superconducting, Kresin1987Gap, Kresin1987Tc, Combescot1989Spectral, Combescot1989Tc}. Second, since bosons are slow, the conventional Migdal-Eliashberg theory focuses on the fermionic sector of the theory~\cite{Bennemann, Marsiglio2020Eliashberg, Protter2021}, so that the effective bosonic propagator has no self-energy from a self-consistent account of the electron-boson interaction; instead, the spectral function of bosons is defined empirically. Considerations beyond small $\lambda_E$ are challenging due to the absence of a small parameter. Currently, there is no experimental evidence for the $\sqrt{\lambda}$ growth of $T_c$ in the adiabatic regime of the strong-coupling limit, raising the question of the validity of Migdal-Eliashberg theory for $\lambda \to \infty$~\cite{Alexandrov2001Breakdown, Esterlis2018Breakdown, Chubukov2020Eliashberg, Yuzbashyan2022, Zhang2024Applicability}. 

In addition to describing superconductivity in conventional metals, Migdal-Eliashberg theory appears as an effective theory for strongly correlated electrons near a quantum critical point, where the order parameter is a scalar boson coupled to electrons; see Refs.~\cite{Millis1992Nearly, Rech2006Quantum, Zhang2024Applicability} and additional references in Ref.~\cite{Zhang2024Applicability}. Theoretical considerations of an electronic Fermi surface coupled to scalar bosons have qualitatively reproduced certain optical and thermodynamic properties of strange metals~\cite{Phillips2022} (an unconventional metallic state above the critical temperature in high-$T_c$ superconductors without long-lived quasiparticles due to strong electronic correlations, which thus resides outside the Fermi-liquid paradigm) from a two-dimensional ($2$d) variant of the Yukawa-Sachdev-Ye-Kitaev (Y-SYK) model~\cite{Patel2023Universal, Li2024Strange}. Below, we focus on this class of systems.

The Y-SYK model~\cite{Esterlis2019Cooper, Wang2020Solvable} describes electrons randomly coupled to bosonic modes and has become instrumental in theoretical studies of correlated electron-boson systems, ranging from thermodynamics~\cite{Wang2020Quantum, Wang2021Phase, Pan2021Yukawa, Esterlis2021Large, Patel2023Universal} and transport~\cite{Guo2022Large, Guo2024Cyclotron, Patel2023Universal, Wang2025Linear} to properties of the pairing phase~\cite{Esterlis2019Cooper, Wang2020Solvable, Hauck2020Eliashberg, Classen2021Superconductivity, Choi2022Pairing, Valentinis2023BCS, Valentinis2023Correlation, Li2024Strange,Esterlis2025QCETh}. This model is solvable in the large-$N$ limit, which allows one to self-consistently incorporate the bosonic self-energy within Migdal-Eliashberg theory and, thus, to consider arbitrarily large values of the electron-boson coupling $\lambda$ without a small parameter. Calculations of $T_c$ in some variants of the Y-SYK model have predicted that, for large coupling, $T_c$ saturates to a constant value defined by a bosonic energy scale~\cite{Esterlis2019Cooper, Valentinis2023BCS, Valentinis2023Correlation}, in sharp contrast to the $\sqrt{\lambda}$ behavior~\cite{Allen1975Transition, Allen1983Theory, Kresin1984Superconducting, Kresin1987Gap, Kresin1987Tc, Combescot1989Spectral, Combescot1989Tc} which is absent. However, the studies of $T_c$ as a function of $\lambda$~\cite{Esterlis2019Cooper, Valentinis2023BCS, Valentinis2023Correlation} rely on the absence of spatial structure in the model's large-$N$ action and consider the system as a quantum dot where electrons and bosons have no dispersion. 
At the same time, recent efforts to capture the physics of strange metals within the Y-SYK model in $2$d~\cite{Patel2023Universal, Li2024Strange} are grounded in having a Fermi surface coupled to bosons, where considering dispersive particles is essential to reproducing strange metal observables. Understanding the behavior of $T_c$ in such systems, beyond the weak-coupling limit, is crucial to connecting the physics of the Y-SYK model to superconducting phenomena in non-Fermi liquids and strange metals. 

In this Letter, we self-consistently find $T_c$ for arbitrary $\lambda_E$ for an electron-boson superconductor described by the $2$d Y-SYK model. We demonstrate that in $2$d, there is a crossover between the $\sqrt{\lambda}$ growth of $T_c$ and the saturation of $T_c$. The saturated value of $T_c$ is a fraction of $\varepsilon_F$ and naturally implies an upper bound on $T_c$. This finding qualitatively agrees with the $T_c$ bound reported for a $2$d system~\cite{Hazra2019Bounds} and differs from previous studies of the Y-SYK model~\cite{Esterlis2019Cooper, Valentinis2023BCS, Valentinis2023Correlation}. The two distinct behaviors of $T_c$ are separated by the electron-boson coupling scale $\lambda\sim\lambda_s$, where $\lambda_s = (\varepsilon_F/\omega_D)^2/(2\pi)$ and $\omega_D/\varepsilon_F \ll 1$. For strong electron-boson coupling $1\ll \lambda \ll \lambda_s$, the bosonic self-energy is negligible. As a result, $T_{c}\approx 0.183\,\omega_D \sqrt{\lambda}$ as in the conventional Migdal-Eliashberg theory~\cite{Allen1975Transition, Carbotte1990}. Increasing $\lambda$ beyond $\lambda_s$ leads to saturation of $T_c$ to a universal value of $T_c \approx 0.04 \varepsilon_F$. This saturation effect occurs because the bosonic self-energy becomes dominant for large $\lambda_E$. For a smaller value of the adiabaticity ratio $\omega_D/\varepsilon_F$, a larger coupling is needed to reach the upper bound on $T_c$. Whereas asymptotically strong coupling typically implies $\lambda \to \infty$ and $\omega_D/\varepsilon_F \to 0$, which leads to the $\sqrt{\lambda}$ divergence, the account of the bosonic self-energy bounds the growth of $T_c$ from above for $\lambda_E \to \infty$. In the following analysis, we refer to the latter limit as the asymptotically strong coupling limit. 

We argue that the crossover between the strong and asymptotically strong coupling behaviors of $T_c$ is present for a class of large-$N$ strongly coupled electron-boson superconductors. The $T_c$ results for $1 \ll \lambda \ll \lambda_s$ and $\lambda_E\to\infty$ stem from the negligible and dominant roles, respectively, of the bosonic self-energy in the corresponding regimes, which are inherent in the general structure of the self-consistent Migdal-Eliashberg theory equations in the large-$N$ limit. 
In addition, here we show that tuning the system to distinct coupling regimes within one (Y-SYK) model leads to similar behavior for $T_c$ as in the pairing instability of a quantum critical metal described by different variants of the $\gamma$-model~\cite{Abanov2020gamma01,Wu2020gamma01T,Wu2020gamma1,Wu2021gamma2,Kiessling2025Bounds}. Lastly, on the basis of our analysis, we estimate an upper bound for $T_c$ in cuprates.

{\it The model.---}
We consider the $2$d Y-SYK model describing $N$ non-relativistic electrons $\psi_{i\sigma}$ with quadratic dispersion $\xi_{\bf k}$ and infinite bandwidth~\cite{Patel2018} coupled to $N$ massless bosons $\phi_i$ with the linear dispersion $c q$, where $c$ is the speed of sound. We introduce an upper cutoff for the bosonic momentum $\omega_D/c$, where $\omega_D$ is an analog of the Debye frequency.
The Hamiltonian reads $H = H_\psi + H_\phi + V_{\psi\phi}$, where
\begin{align}
H_\psi &= \sum_{i=1}^N \sum_{\sigma=\uparrow\downarrow}  \int\!  \xi_{\bf k} \psi^\dag_{i\sigma}({\bf k}) \psi_{i\sigma}({\bf k})\frac{d^{2}k}{(2\pi)^2}, \\
H_\phi &= \frac{c^2}{2} \sum_{i=1}^N \int\!  q^2 \phi_i({\bf q}) \phi_i(-{\bf q}) \frac{d^2 q}{(2\pi)^2}, \\
V_{\psi \phi} &= \frac{1}{N} \sum_{ijl=1}^N \sum_{\sigma=\uparrow\downarrow} \int \!  g_{ijl}({\bf r}) \psi^\dag_{i\sigma}({\bf r}) \psi_{j\sigma}({\bf r}) \phi_l({\bf r}) {d^2 r}, \label{V}
\end{align}
where $i=1,\cdots,N$ labels fermions and bosons, and $\sigma=\uparrow\downarrow$ is the electron's spin. 
The electron-boson coupling constants $g_{ijl}$ are sampled from a real random Gaussian distribution~\footnote{Sampling the coupling constants from the real random Gaussian distribution (instead of the complex one) ensures the presence of anomalous averages below $T_c$ in the large-$N$ theory \cite{Esterlis2019Cooper}} with zero mean and finite variance, $\langle g_{ijl}({\bf r}) g_{ijl} ({\bf r}')\rangle = 2 g^2 \delta({\bf r}-{\bf r}')$, so that the coupling is spatially local~\cite{Li2024Strange, Guo2024Cyclotron}. 

We perform the disorder average for the model and then decouple the interaction using time-nonlocal Lagrange multipliers~\cite{Esterlis2019Cooper}. In the large-$N$ limit, and in the metallic regime when the anomalous components of the electronic Green's function vanish, we obtain the following system of self-consistent equations~\cite{SupMat}:
\begin{align}
\Sigma(i\omega_n) &= g^2 T \sum_{\omega_{n'}} {\cal D}(i\omega_n-i\omega_{n'})  {\cal G}(i\omega_{n'}), \label{Sigma_ME} \\ 
{\cal G}(i\omega_n) &= \int\!  \frac{1}{i\omega_n -\xi_{\bf k}-\Sigma(i\omega_n)}  \frac{d^{2}k}{(2\pi)^2},  \label{G_ME} \\
\Pi(i\Omega_m) &= -2g^2 T \sum_{\omega_n} {\cal G}(i\omega_n) {\cal G}(i\omega_n+i\Omega_m), \label{Pi_ME} \\ 
{\cal D}(i\Omega_m) &=\int_0^{\omega_D/c} \! \frac{1}{\Omega_m^2 + c^2 q^2 - \Pi(i\Omega_m)} \frac{qdq}{2\pi}. \label{D_ME}
\end{align}
The momentum-resolved propagators for fermions and bosons are $G(i\omega_n, {\bf k}) = 1/(i\omega_n -\xi_{\bf k}-\Sigma(i\omega_n))$ and $D(i\Omega_m, {\bf q}) = 1/(\Omega_m^2 + c^2 q^2 - \Pi(i\Omega_m))$. The frequencies $\omega_n = (2n+1)\pi T$ and $\Omega_m = 2m\pi T$ are fermionic and bosonic Matsubara frequencies at temperature $T$. Equations (\ref{Sigma_ME}-\ref{D_ME}) are solvable for arbitrary values of $g^2$. 

The locality of the electron-boson coupling leads to factorization of the momentum integrals, making the self-energies \eqref{Sigma_ME} and \eqref{Pi_ME} momentum independent. This allows one to exactly compute the effective propagators ${\cal G}$ and ${\cal D}$ for fermions and bosons. To do so, it is convenient to express the electronic self-energy as $\Sigma(i\omega_n) = i\omega_n(1-Z(i\omega_n))$, where $Z(i\omega_n)$ is the even-frequency renormalization function that incorporates the effects of the interaction by renormalizing the $i\omega_n$ term in the electronic Green's function~\cite{AGD}. The poles of the integrand in Eq.~(\ref{G_ME}), given by $\xi_{\bf k} = i\omega_n Z(i\omega_n)$, lie on the same side of the complex plane as in the $Z=1$ non-interacting case. Evaluating Eq.~\eqref{G_ME}, we obtain the interaction-independent effective electronic Green's function ${\cal G}(i\omega_n) = -i\pi \nu {\rm sgn}(\omega_n)$, where $\nu = m_0/(2\pi)$ is the density of states of the electrons per spin in $2$d and $m_0$ is the mass of the electron. Substituting this result into the self-energy of the boson (\ref{Pi_ME}), we find $\Pi(i\Omega_m)=-2\pi\nu^2 g^2 |\Omega_m|$. Using Eq.~(\ref{D_ME}), we then obtain the effective bosonic propagator ${\cal D}(i\Omega_m) = L(i\Omega_m)/(4\pi c^2)$, where 
\begin{equation}
L(i\Omega_m)= \log \!\left[1\!+\!\frac{1}{(\frac{\Omega_m}{\omega_D})^2 \! + \! 2 \pi \lambda\frac{|\Omega_m|}{\varepsilon_F}} \right]. 
\label{L_eff}
\end{equation}
The dimensionless electron-boson coupling constant is $\lambda=\nu g^2/(4\pi c^2)$. We assume that the Debye momentum $k_D$ is of the same order as the Fermi momentum $k_F$~\cite{AGD, Abrikosov1988Fundamentals}. In the following, we refer to $L(i\Omega_m)$ as the bosonic propagator.

We obtain $T_c$, implying the temperature of pair formation~\cite{Chubukov2005Superconductivity}, from the linearized equation for the superconducting order parameter $\Phi$:
\begin{align}
\Phi(i\omega_n) &= g^2 T_c \sum_{\omega_{n'}} {\cal D}(i\omega_n-i\omega_{n'})  {\cal F}(i\omega_{n'}), \label{Phi} \\
{\cal F}(i\omega_n) &= \!\int\! G(i\omega_n, {\bf k})G(-i\omega_n, -{\bf k}) \Phi(i\omega_n)\frac{d^{2}k}{(2\pi)^2}. \label{F_eff}
\end{align}
The effective linearized anomalous component of Gor'kov's Green's function ${\cal F}$ employs factorization of the momentum integral, similarly to ${\cal G}$ in the normal-state self-energy (\ref{Sigma_ME}). To obtain the linearized gap equation, we (i) substitute Eq.~(\ref{F_eff}) into Eq.~(\ref{Phi}); (ii) evaluate $\!\int\! G(i\omega_n, {\bf k})G(-i\omega_n, -{\bf k})\frac{d^{2}k}{(2\pi)^2} = \nu \pi/(|\omega_n| Z(i\omega_n))$; (iii) define  $\Delta(i\omega_n) = \Phi(i\omega_n)/Z(i\omega_n)$;  (iv) express the renormalization function via the fermionic self-energy, $Z(i\omega_n)=1+i\Sigma(i\omega_n)/\omega_n$, using $\cal G$ and $\cal D$; (v) move $L$-dependent terms to the right-hand side~\cite{SupMat}:
\begin{equation}
\Delta(i\omega_n) \!=\! \pi\!\lambda  T_{\!c} \!\!\sum_{\omega_{n'\!}}\!\! \bigg[\!\Delta(i\omega_{n'}\!) -\!\frac{\omega_{n'\!}\!}{\omega_n} \!\Delta(i\omega_n) \!\bigg]\!
\frac{L(i\omega_n\!\!-\!i\omega_{n'}\!)}{|\omega_{n'}\!|}. \label{T_c_eq_1}
\end{equation}
The resonant term ($n=n'$) does not contribute to $T_c$, as in the conventional Migdal-Eliashberg theory~\cite{Chubukov2020Eliashberg}. Excluding this term, the linearized gap equation for positive frequencies becomes 
\begin{align} \nonumber
&z(i\omega_n) \Delta(i\omega_n) = \pi \lambda T_c \sum_{n'=0}^{+\infty} \frac{\Delta(i\omega_{n'})}{\omega_{n'}} \\
& \times \bigg[(1\!-\!\delta_{nn'}) L(i\omega_n\!-\!i\omega_{n'}) \!+\! L(i\omega_n\!+\!i\omega_{n'})  \bigg], \label{T_c_eq_2}
\end{align}
where the new renormalization function $z(i\omega_n)= 1 + (2\pi\lambda  T_c/\omega_n) \sum_{m=1}^n L(i\Omega_m)$ is analogous to $Z(i\omega_n)$ but excludes the resonant term.

Equation~\eqref{T_c_eq_2} is obtained non perturbatively (in $\lambda$), and thus it determines $T_c$ for arbitrary values of electron-boson coupling.
Since Eq.~(\ref{T_c_eq_2}) is a linear equation for $\Delta$, $T_c$ can be found from a zero of the determinant~\cite{Kresin1987Tc}, 
\begin{equation}
\det[{\cal M}-{\cal I}] = 0, \label{det}
\end{equation}
where ${\cal I}$ is the identity matrix in the space of positive frequencies and ${\cal M}$ is the matrix with elements given by
\begin{equation}
{\cal M}_{nn'} \!=\! \pi\!\lambda T_c \frac{(1\!-\!\delta_{nn'}) L(i\omega_n\!-\!i\omega_{n'}\!) \!+\! L(i\omega_n\!+\!i\omega_{n'}\!)}{z(i\omega_n)\omega_{n'}}. \label{M}
\end{equation}
This allows us to compute $T_c$ for any value of $\lambda$. In what follows, we compute $T_c$ for asymptotically strong, strong, and weak coupling numerically and analytically from Eqs.~\eqref{L_eff},\eqref{det},\eqref{M}. The numerical solution for $T_c$ uses $100$ Matsubara frequencies except for the regime $\omega_D/\varepsilon_F\geq 1$ and $\lambda< 1$, where we use $300$ frequencies~\cite{SupMat}. 

\begin{figure*}[t!]
\center
\includegraphics[width=1.\linewidth]{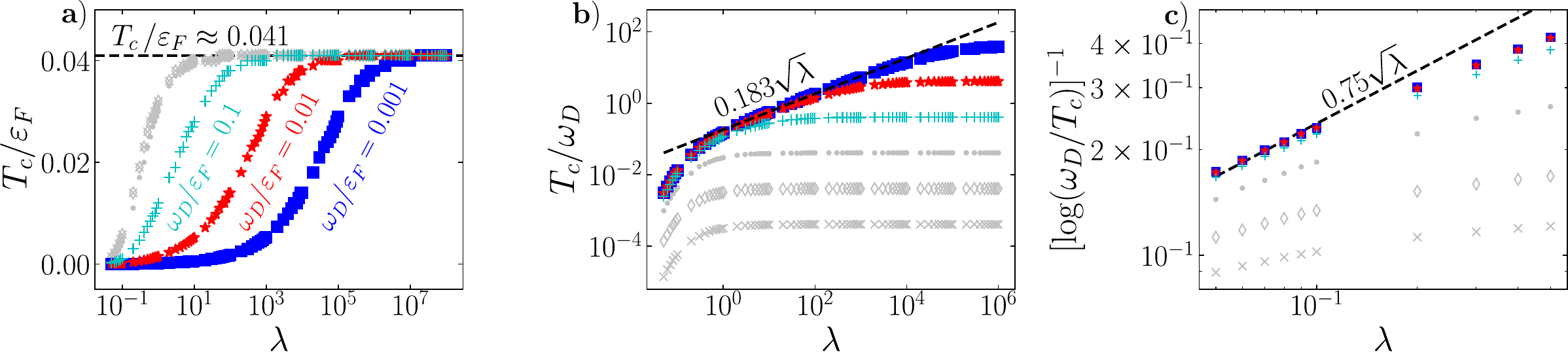}  
\caption{\small \label{fig:Tc} {\bf The critical temperature} ($T_c$) {\bf as a function of the electron-boson coupling strength} ($\lambda$). {\bf a}) $T_c$ in units of the Fermi energy $\varepsilon_F$ over a large range of $\lambda$ for several values of $\omega_D/\varepsilon_F$. The legend for the data points is the same throughout all panels: blue squares, red stars, and cyan pluses show $\omega_D/\varepsilon_F=0.001$, $0.01$, and $0.1$ data, respectively. The data marked by gray dots, diamonds, and crosses show $T_c$ for $\omega_D/\varepsilon_F = 1$, $10$, and $100$. The dashed horizontal line shows the saturated $T_c$ in the asymptotically strong coupling regime. {\bf b}) $T_c$ in units of the Debye frequency $\omega_D$ as a function of $\lambda$ on a log-log scale. The dashed black line shows the $\sqrt{\lambda}$ increase of $T_c$ at strong coupling. {\bf c}) Inverse logarithm of $\omega_D/T_c$ on a log-log scale at weak coupling. The dashed black line is a fit for the six left points of the $\omega_D/\varepsilon_F=0.001$ dataset.}
\end{figure*}

{\it Asymptotically strong coupling.---} 
For asymptotically strong coupling $\lambda_E \to \infty$, we expand the effective bosonic propagator in $1/\lambda_E$:
\begin{equation}
L(i\Omega_m) \!=\! \log \!\left[1\!+\!\frac{1}{(\frac{\Omega_m}{\omega_D})^2 \! + \! 2\pi\! \lambda_E\frac{|\Omega_m|}{\omega_D}} \right] \!\simeq\!\frac{\varepsilon_F}{2\pi\!\lambda |\Omega_m|}. \label{L_asc}
\end{equation}
The pairing problem (\ref{T_c_eq_1}) with the bosonic propagator (\ref{L_asc}) maps to the case of the $\gamma$-model with $\gamma=1$, which describes the low-energy physics of a quantum critical metal via the effective dynamical electron-electron interaction~\cite{Wu2020gamma1}, with the effective coupling constant $\varepsilon_F/(2\pi)$ independent of $\lambda$ and $\omega_D$.

From Eq.~(\ref{L_asc}), the characteristic equation (\ref{det}) becomes independent of the electron-boson coupling $\lambda$ and UV bosonic momentum cutoff $\omega_D$:
\begin{equation}
{\cal M}_{nn'} = \frac{1}{2}  \!\frac{\varepsilon_F T_c }{z(i\omega_n) \omega_{n'}}  \bigg[\frac{1\!-\!\delta_{nn'}}{|\omega_n\!-\!\omega_{n'}|} \!+\! \frac{1}{\omega_n\!+\!\omega_{n'}}  \bigg], \label{M_asc}
\end{equation}
where $z(i\omega_n) = 1+ \varepsilon_F H_n/(2\pi\omega_n)$, and $H_n$ is the $n$-th harmonic number. 

To estimate $T_c$ analytically, it suffices to consider the first three Matsubara frequencies in the characteristic equation (\ref{det}) for the matrix (\ref{M_asc}), which leads to $T_c \approx 0.038 \varepsilon_F$. In Fig.~\ref{fig:Tc}{\bf a}, we plot $T_c$ (in units of $\varepsilon_F$) as numerically determined from Eq.~\eqref{det} using the exact bosonic propagator \eqref{L_eff}. In the regime of $\lambda_E\to\infty$, $T_c$ saturates to a universal value $T_c\approx 0.041\varepsilon_F$, independent of the coupling strength and $\omega_D$, providing an upper bound on $T_c$. The saturated value is very close to the three-frequencies approximation \footnote{The number $0.04$ agrees with the one predicted in the $\gamma=1$ model: $T_c \approx 0.25 {\bar g}$~\cite{Wu2020gamma1}, when accounting for the effective coupling constant $\bar{g} = \varepsilon_F/(2\pi)$}. 
Saturation of the upper bound does not require a small adiabatic ratio. However, for smaller $\omega_D/\varepsilon_F$, larger $\lambda$ is needed to reach the upper bound on $T_c$. In the opposite case, when the speed of sound is greater than the Fermi velocity, $T_c$ begins to saturate at $\lambda\sim 10$, independent of the particular value of $\omega_D/\varepsilon_F$ once $\omega_D/\varepsilon_F\gtrsim 1$.  

The emergence of a critical temperature bound for the asymptotically strong coupling $\lambda_E\to \infty$:
\begin{equation}
T_c \approx 0.04 \, \varepsilon_F,
\label{T_c_asc}
\end{equation}
qualitatively agrees with the upper bound on $T_c$ prediction for a $2$d system~\cite{Hazra2019Bounds}. This result remains valid beyond the massless boson considered here. Indeed, the large-$\lambda_E$ expansion (\ref{L_asc}) holds for the massive case too, since the boson mass $\epsilon$, if present in Eq.~(\ref{L_eff}), would be negligible compared to its self-energy~\cite{Chubukov2005Superconductivity}. 

{\it Strong coupling.---}
Studies of $T_c$ for strong coupling, $\lambda \gg 1$, within Migdal-Eliashberg theory generally rely on $\omega_D/\varepsilon_F \ll 1$. 
We expand the bosonic propagator (\ref{L_eff}) in $\omega_D/\varepsilon_F$,
\begin{equation} 
L(i\Omega_m) \!=\! \log \!\left[1\!+\!\frac{(\frac{\omega_D}{\varepsilon_F})^2}{(\frac{\Omega_m}{\varepsilon_F})^2 \! + \! 2\pi \lambda (\frac{\omega_D}{\varepsilon_F})^2 \frac{|\Omega_m|}{\varepsilon_F}} \right]  \!\simeq\! \frac{\omega_D^2}{\Omega_m^2}, \label{L_sc}
\end{equation}
which is valid for coupling $\lambda \ll \lambda_s = (\varepsilon_F/\omega_D)^2/(2\pi)$. Consequently, the strong-coupling regime is restricted to $1 \ll \lambda \ll \lambda_s$. The expansion of the bosonic propagator \eqref{L_sc} naturally implies the limit $\omega_D/\varepsilon_F \to 0$, which is often assumed for electron-boson superconductivity and neglects the self-energy of the boson. Accordingly, the propagator \eqref{L_sc} matches the one in the massless limit for electron-boson superconductors at strong coupling $\lambda\gg 1$ in conventional Eliashberg theory~\cite{Marsiglio1991Gap}, and defines the pairing problem \eqref{T_c_eq_1} similar to the one for the $\gamma$-model with $\gamma=2$ and the effective coupling constant $\omega_D\sqrt{\lambda}$~\cite{Wu2021gamma2}.

The pairing problem simplifies to
\begin{equation}
   {\cal M}_{nn'} \!=\! \pi\!\lambda\omega_D^2  \frac{T_c}{z(i\omega_n) \omega_{n'}}  \left[\!\frac{1\!-\!\delta_{nn'}}{|\omega_n\!-\!\omega_{n'}|^2} \!+\! \frac{1}{(\omega_n\!+\!\omega_{n'}\!)^2}  \!\right] \label{M_sc} 
\end{equation}
and $z(i\omega_n) = 1+ \lambda\omega_D^2 H_{n,2}/(2\pi T_c \omega_n)$, which are independent of $\varepsilon_F$, and $H_{n,2}$ is the $n$-th harmonic number of the second order. As in the asymptotically strong coupling regime, the solution of the $T_c$ equation (\ref{det}) quickly converges, and accounting for the first four Matsubara frequencies in the matrix (\ref{M_sc}) is sufficient to recover the seminal result for $T_c$ in the strong-coupling limit~\cite{Allen1975Transition, Allen1983Theory, Kresin1984Superconducting, Kresin1987Gap, Kresin1987Tc, Combescot1989Spectral, Combescot1989Tc}:
\begin{equation}
T_c \approx 0.183 \, \omega_D \sqrt{\lambda}. \label{T_c_sc}
\end{equation}
Inclusion of the bosonic mass does not change $T_c$, since the expansion (\ref{L_sc}) is valid as long as the boson mass is negligible compared to $\varepsilon_F$ or $2\pi T$ and leads to the same result (\ref{T_c_sc})~\cite{Carbotte1990}. 

The critical temperature obtained in Eq.~(\ref{T_c_sc}) agrees with the known behavior of strongly coupled electron-boson superconductors~\cite{Carbotte1990}, and illustrates the $T_c$ divergence for $\lambda \gg 1$. In the absence of the bosonic self-energy in Eq.~\eqref{L_sc}, as often appears in Migdal-Eliashberg theory, $T_c$ at strong coupling (\ref{T_c_sc}) continues to rise without saturation with increasing $\lambda$~\footnote{For comparison, we demonstrate the unconstrained growth of $T_c$ in the strong-coupling limit of Migdal-Eliashberg theory in the absence of the self-energy of the boson in the Supplemental Material}, e.g., see Ref.~\cite{Heath2024}. However, a self-consistent account of the self-energy of the boson for $\lambda_E\to\infty$ restricts the $\sqrt{\lambda}$ growth of $T_c$ and provides an upper bound on $T_{c}$ (\ref{T_c_asc}) from which it follows that there is no divergence. The $T_c$ result for the asymptotically strong coupling is non-perturbative. It accounts for the bosonic self-energy for arbitrarily large values of the electron-boson coupling. In turn, achieving the saturated value of $T_c$ requires a larger $\lambda$ for a smaller $\omega_D/\varepsilon_F$. Eventually, the $\sqrt{\lambda}$ growth of $T_c$ saturates, provided $\lambda$ is large enough. In fact, these two regimes of $T_c$ behavior are separated by $\lambda\sim\lambda_s$, which is justified by the validity of the expansion of the bosonic propagator (\ref{L_sc}).

In Fig.~\ref{fig:Tc}{\bf b}, we show the numerical solution of the $T_c$ equation~\eqref{det} with the exact bosonic propagator (\ref{L_eff}), where $T_c$ is plotted in units of the Debye frequency for different values of $\omega_D/\varepsilon_F$ as a function of the electron-boson coupling $\lambda$. As we argued in Eq.~\eqref{T_c_sc} for small values of $\omega_D$, $T_{c}\approx0.183\,\omega_{D}\sqrt{\lambda}$ for $\lambda \gg 1$ until the coupling approaches the value of $\lambda \sim \lambda_s$ where the crossover to saturation occurs. For $\omega_D\gtrsim\varepsilon_F$, the Allen-Dynes~\cite{Allen1975Transition} regime of $T_c$ (\ref{T_c_sc}) is absent.

{\it The crossover.---}
The occurrence of the crossover in $T_c$ at strong/asymptotically strong coupling relies on the role of the bosonic self-energy in Migdal-Eliashberg theory: the bosonic self-energy is negligible for the strong coupling under $\omega_D/\varepsilon_F\ll 1$, leading to $\sqrt{\lambda}$ scaling of $T_c$; the bosonic self-energy becomes dominant for the asymptotically strong coupling and results in saturation of $T_c$. Once we associate asymptotically strong coupling with the primary role of the bosonic self-energy, the saturation of $T_c$ can be qualitatively argued from the self-consistent form of Migdal-Eliashberg theory. Indeed, Eqs.~(\ref{Sigma_ME}-\ref{D_ME}) exemplify the summable and self-consistent version of the Migdal theory for electron-boson systems~\cite{Migdal1958}, where the approximation on the vertex part $\Gamma=\Gamma_0$ (in our case, $\Gamma_0=g^2$) is exact and, consequently, the bosonic self-energy can be accounted for at all couplings. Equations~(\ref{Phi},\ref{F_eff}) then determine $T_c$. Taking $g \to \infty$ (which should be consistent with $\lambda_E\to\infty$ for the appropriately defined electron-boson coupling constant and adiabaticity ratio), we consider the bosonic propagator to be primarily defined by its self-energy (\ref{Pi_ME}) as $-1/\Pi$. Substituting the bosonic propagator into $\Sigma$ (\ref{Sigma_ME}) and $\Phi$ (\ref{Phi}), we notice that $g^2$ cancels, making the pairing problem independent of the electron-boson coupling, similar to the cancellation of $\lambda$ in Eq.~(\ref{M_asc}). The particular Y-SYK model considered here is convenient for the analytical calculations, though not essential for the qualitative argument above. Hence, we expect that self-consistency in Migdal-Eliashberg theory should lead to saturation of $T_c$ in the asymptotically strong coupling limit, provided that vertex corrections are suppressed in this limit as well. The latter scenario is the case for the class of large-$N$ theories (containing multiple types of random couplings) and aims to describe strange-metal physics~\cite{Patel2023Universal,Li2024Strange}, where the vertex corrections are negligible in the large-$N$ limit.

{\it Weak coupling.---}
It is not possible to expand the logarithm in the bosonic propagator in Eq.~\eqref{L_eff} for weak coupling in a manner similar to that in Eqs.~\eqref{L_asc}  and \eqref{L_sc}, since $T_c$ is typically exponentially suppressed in $\lambda$~\cite{McMillan1968, Karakozov1975}. The low temperatures in the Matsubara frequencies in the bosonic propagator do not allow using $\omega_D^2$ as a small parameter in an expansion, unless there is a boson mass. Hence, preserving the logarithmic form of the bosonic propagator is crucial for the massless case. We can still state the pairing problem in the lowest order in $\lambda$ and $\omega_D/\varepsilon_F$, which takes the form of Eq.~\eqref{T_c_eq_2} with $z(i\omega_n) \simeq 1$ and $L(i\Omega_m) \simeq \log (1+\omega_D^2/\Omega_m^2)$, where the one appearing in the bosonic propagator ensures convergence of the Matsubara sum. In the low-temperature limit, the main contribution to the sum in Eq.~\eqref{T_c_eq_2} arises from the lowest Matsubara frequency. Consequently, we approximate $L(i\omega_n\pm i\omega_{n'})$ by $L(i\omega_n\pm i\omega_0)$ in the pairing problem. The latter approximation simplifies the $T_c$ equation to
\begin{align} \nonumber
    &\Delta(i\omega_n) = \pi \lambda T_c \sum_{n'=0}^{+\infty} \frac{\Delta(i\omega_{n'})}{\omega_{n'}} \\
    & \times \bigg[(1\!-\!\delta_{n0}) L(i\omega_n\!-\!i\omega_0) \!+\! L(i\omega_n\!+\!i\omega_0)  \bigg], \label{T_c_eq_wc_1}
\end{align}
by factorizing the sum over $n'$. Next, we multiply Eq.~\eqref{T_c_eq_wc_1} by $1/\omega_n$, sum the resulting equation over $n$, and take the continuous limit $\omega_n \to \omega$. Using the fact that in the low-temperature limit $L(i\omega-i\omega_0) + L(i\omega+i\omega_0)\simeq 2 L(i\omega)$, we obtain the $T_c$ equation for weak coupling: $1 = \lambda L(2i\omega_0) +\lambda \int_{\omega_1}^{+\infty}  d\omega \,L(i\omega)/\omega$, where $\omega_0 = \pi T_c$ and $\omega_1 = 3\pi T_c$ contain the dependence on $T_c$. From here, the main low-temperature contribution gives: $\log(\omega_D/T_c)  = 1/\sqrt{\lambda}$, reminiscent of color superconductivity~\cite{Son1999Superconductivity, Wilczek1999, Pisarski2000, Chubukov2005Superconductivity}. This type of $\lambda$ scaling of $T_c$ appears for electron-boson systems with a massless boson at weak coupling~\cite{Chubukov2005Superconductivity}, and for systems where a pairing instability occurs from the non-Fermi-liquid state~\cite{Metlitski2015Cooper, Gnezdilov2022Solvable}, in analogy with the $\gamma=0^{+}$ limit of the $\gamma$-model of critical electrons~\cite{Abanov2020gamma01, Wu2020gamma01T}.

In Fig.~\ref{fig:Tc}{\bf c}, we show the numerical solution of the pairing problem \eqref{det} with the exact bosonic propagator (\ref{L_eff}) in the weak-coupling regime. We plot $[\log(\omega_D/T_c)]^{-1}$ versus $\lambda$ on a log-log scale to distinguish the power-law behavior. The $\omega_{D}/\varepsilon_F \ll 1$ curves merge for smaller values of $\lambda$. The smaller the adiabaticity ratio, the closer the curves lie. In contrast, if we measure $T_c$ in units of $\varepsilon_F$, the $T_c/\varepsilon_F$ curves will merge for $\omega_D/\varepsilon_F \gg 1$~\cite{SupMat}. Fitting the first six data points in Fig.~\ref{fig:Tc}{\bf c} for $\omega_D/\varepsilon_F=0.001$ with $\alpha\sqrt{\lambda}$, where $\alpha$ is the fitting parameter, we get $\left[\log(\omega_D/T_c)\right]^{-1} = 0.75 \sqrt{\lambda}$. The result of the fit, $\alpha=0.75\pm 0.01$, is of the same order of magnitude as the analytical estimate that gives $\alpha = 1$. 

{\it Conclusion.---}
Apart from electron-phonon systems, Migdal-Eliashberg theory also applies to correlated electronic systems near quantum criticality, where the effective electron-boson interaction emerges from the coupling of the electrons to an order parameter that behaves as a scalar boson~\cite{Zhang2024Applicability}. By modeling the electron-boson interaction with the $2$d Y-SYK model, it is possible to qualitatively address certain properties of strange metals~\cite{Patel2023Universal, Li2024Strange}, including the properties of the superconducting state~\cite{Li2024Strange}. The $2$d Y-SYK model is a self-consistent, solvable variant of Migdal-Eliashberg theory which is valid for arbitrary electron-boson coupling due to the absence of vertex corrections. Based on these attributes, we computed $T_c$ for an electron-boson superconductor for an unconstrained coupling strength. We found that the $T_c$ behavior recovers the pairing instability of quantum critical metals. The weak, strong, and asymptotically strong coupling regimes in our model correspond to $\gamma=0^{+},2,$ and $1$, respectively, in the $\gamma$-model of critical electronic systems~\cite{Abanov2020gamma01, Wu2020gamma1,Wu2021gamma2} with appropriately renormalized coupling constants. We showed that the self-consistent account of the bosonic self-energy, neglected in the electron-phonon Migdal-Eliashberg theory, leads to a crossover of $T_c$ in the strong-coupling regime. For strong coupling $1\ll \lambda\ll \lambda_s$, $T_c$ increases as $T_c \approx 0.183 \, \omega_D\sqrt{\lambda}$, as predicted in the conventional Migdal-Eliashberg theory in the adiabatic regime. For asymptotically strong coupling $\lambda_E \to \infty$, $T_c$ saturates to a constant value independent of the electron-boson coupling strength. For the $2$d Y-SYK model, we have $T_c \approx 0.04 \varepsilon_F$ independent of the bosonic momentum cutoff $\omega_D$ and $\lambda$. The coupling strength needed to reach the saturated value depends on the adiabaticity ratio. We expect our results to hold qualitatively for electron-boson systems if the vertex corrections in Migdal-Eliashberg theory are negligible. Based on our analysis, we can estimate the magnitude of the upper bound on $T_c$ for LSCO at doping $x=0.24$, where the Fermi level is at $-253\, {\rm meV}$~\cite{Michon2023Reconciling}, to be $0.04 \, \varepsilon_F \approx 117\, {\rm K}$, which is indeed higher than the values~\cite{Badoux2016} of $T_c$ measured in a smaller range of $x$ values. Our upper bound is also below the bound $T_c \leq\varepsilon_F/8$ estimated for $2$d systems~\cite{Hazra2019Bounds}.

{\it Acknowledgment.---} We thank Andrey V. Chubukov for insightful comments on our results. This research was supported by funding provided by Dartmouth College.

\bibliography{refs}

\begin{thebibliography}{11}%
\makeatletter
\providecommand \@ifxundefined [1]{%
 \@ifx{#1\undefined}
}%
\providecommand \@ifnum [1]{%
 \ifnum #1\expandafter \@firstoftwo
 \else \expandafter \@secondoftwo
 \fi
}%
\providecommand \@ifx [1]{%
 \ifx #1\expandafter \@firstoftwo
 \else \expandafter \@secondoftwo
 \fi
}%
\providecommand \natexlab [1]{#1}%
\providecommand \enquote  [1]{``#1''}%
\providecommand \bibnamefont  [1]{#1}%
\providecommand \bibfnamefont [1]{#1}%
\providecommand \citenamefont [1]{#1}%
\providecommand \href@noop [0]{\@secondoftwo}%
\providecommand \href [0]{\begingroup \@sanitize@url \@href}%
\providecommand \@href[1]{\@@startlink{#1}\@@href}%
\providecommand \@@href[1]{\endgroup#1\@@endlink}%
\providecommand \@sanitize@url [0]{\catcode `\\12\catcode `\$12\catcode
  `\&12\catcode `\#12\catcode `\^12\catcode `\_12\catcode `\%12\relax}%
\providecommand \@@startlink[1]{}%
\providecommand \@@endlink[0]{}%
\providecommand \url  [0]{\begingroup\@sanitize@url \@url }%
\providecommand \@url [1]{\endgroup\@href {#1}{\urlprefix }}%
\providecommand \urlprefix  [0]{URL }%
\providecommand \Eprint [0]{\href }%
\providecommand \doibase [0]{https://doi.org/}%
\providecommand \selectlanguage [0]{\@gobble}%
\providecommand \bibinfo  [0]{\@secondoftwo}%
\providecommand \bibfield  [0]{\@secondoftwo}%
\providecommand \translation [1]{[#1]}%
\providecommand \BibitemOpen [0]{}%
\providecommand \bibitemStop [0]{}%
\providecommand \bibitemNoStop [0]{.\EOS\space}%
\providecommand \EOS [0]{\spacefactor3000\relax}%
\providecommand \BibitemShut  [1]{\csname bibitem#1\endcsname}%
\let\auto@bib@innerbib\@empty
\bibitem [{\citenamefont {Esterlis}\ and\ \citenamefont
  {Schmalian}(2019)}]{Esterlis2019Cooper}%
  \BibitemOpen
  \bibfield  {author} {\bibinfo {author} {\bibfnamefont {I.}~\bibnamefont
  {Esterlis}}\ and\ \bibinfo {author} {\bibfnamefont {J.}~\bibnamefont
  {Schmalian}},\ }\bibfield  {title} {\bibinfo {title} {Cooper pairing of
  incoherent electrons: An electron-phonon version of the {Sachdev-Ye-Kitaev}
  model},\ }\href {https://doi.org/10.1103/PhysRevB.100.115132} {\bibfield
  {journal} {\bibinfo  {journal} {Phys. Rev. B}\ }\textbf {\bibinfo {volume}
  {100}},\ \bibinfo {pages} {115132} (\bibinfo {year} {2019})}\BibitemShut
  {NoStop}%
\bibitem [{\citenamefont {Esterlis}\ \emph {et~al.}(2021)\citenamefont
  {Esterlis}, \citenamefont {Guo}, \citenamefont {Patel},\ and\ \citenamefont
  {Sachdev}}]{Esterlis2021Large}%
  \BibitemOpen
  \bibfield  {author} {\bibinfo {author} {\bibfnamefont {I.}~\bibnamefont
  {Esterlis}}, \bibinfo {author} {\bibfnamefont {H.}~\bibnamefont {Guo}},
  \bibinfo {author} {\bibfnamefont {A.~A.}\ \bibnamefont {Patel}},\ and\
  \bibinfo {author} {\bibfnamefont {S.}~\bibnamefont {Sachdev}},\ }\bibfield
  {title} {\bibinfo {title} {Large-{$N$} theory of critical {F}ermi surfaces},\
  }\href {https://doi.org/10.1103/PhysRevB.103.235129} {\bibfield  {journal}
  {\bibinfo  {journal} {Phys. Rev. B}\ }\textbf {\bibinfo {volume} {103}},\
  \bibinfo {pages} {235129} (\bibinfo {year} {2021})}\BibitemShut {NoStop}%
\bibitem [{\citenamefont {Abrikosov}\ \emph {et~al.}(1965)\citenamefont
  {Abrikosov}, \citenamefont {Gor'kov},\ and\ \citenamefont
  {Dzyaloshinskii}}]{AGD}%
  \BibitemOpen
  \bibfield  {author} {\bibinfo {author} {\bibfnamefont {A.~A.}\ \bibnamefont
  {Abrikosov}}, \bibinfo {author} {\bibfnamefont {L.~P.}\ \bibnamefont
  {Gor'kov}},\ and\ \bibinfo {author} {\bibfnamefont {I.~Y.}\ \bibnamefont
  {Dzyaloshinskii}},\ }\href@noop {} {\emph {\bibinfo {title} {Quantum field
  theoretical methods in statistical physics}}},\ \bibinfo {edition} {2nd}\
  ed.\ (\bibinfo  {publisher} {Pergamon press Ltd., Oxford},\ \bibinfo {year}
  {1965})\BibitemShut {NoStop}%
\bibitem [{\citenamefont {Abrikosov}(1988)}]{Abrikosov1988Fundamentals}%
  \BibitemOpen
  \bibfield  {author} {\bibinfo {author} {\bibfnamefont {A.~A.}\ \bibnamefont
  {Abrikosov}},\ }\href@noop {} {\emph {\bibinfo {title} {Fundamentals of the
  Theory of Metals}}}\ (\bibinfo  {publisher} {North-Holland},\ \bibinfo {year}
  {1988})\BibitemShut {NoStop}%
\bibitem [{\citenamefont {Patel}\ \emph {et~al.}(2018)\citenamefont {Patel},
  \citenamefont {McGreevy}, \citenamefont {Arovas},\ and\ \citenamefont
  {Sachdev}}]{Patel2018}%
  \BibitemOpen
  \bibfield  {author} {\bibinfo {author} {\bibfnamefont {A.~A.}\ \bibnamefont
  {Patel}}, \bibinfo {author} {\bibfnamefont {J.}~\bibnamefont {McGreevy}},
  \bibinfo {author} {\bibfnamefont {D.~P.}\ \bibnamefont {Arovas}},\ and\
  \bibinfo {author} {\bibfnamefont {S.}~\bibnamefont {Sachdev}},\ }\bibfield
  {title} {\bibinfo {title} {Magnetotransport in a model of a disordered
  strange metal},\ }\href {https://doi.org/10.1103/PhysRevX.8.021049}
  {\bibfield  {journal} {\bibinfo  {journal} {Phys. Rev. X}\ }\textbf {\bibinfo
  {volume} {8}},\ \bibinfo {pages} {021049} (\bibinfo {year}
  {2018})}\BibitemShut {NoStop}%
\bibitem [{\citenamefont {Chubukov}\ \emph {et~al.}(2020)\citenamefont
  {Chubukov}, \citenamefont {Abanov}, \citenamefont {Esterlis},\ and\
  \citenamefont {Kivelson}}]{Chubukov2020Eliashberg}%
  \BibitemOpen
  \bibfield  {author} {\bibinfo {author} {\bibfnamefont {A.~V.}\ \bibnamefont
  {Chubukov}}, \bibinfo {author} {\bibfnamefont {A.}~\bibnamefont {Abanov}},
  \bibinfo {author} {\bibfnamefont {I.}~\bibnamefont {Esterlis}},\ and\
  \bibinfo {author} {\bibfnamefont {S.~A.}\ \bibnamefont {Kivelson}},\
  }\bibfield  {title} {\bibinfo {title} {Eliashberg theory of phonon-mediated
  superconductivity — when it is valid and how it breaks down},\ }\href
  {https://doi.org/10.1016/j.aop.2020.168190} {\bibfield  {journal} {\bibinfo
  {journal} {Annals of Physics}\ }\textbf {\bibinfo {volume} {417}},\ \bibinfo
  {pages} {168190} (\bibinfo {year} {2020})}\BibitemShut {NoStop}%
\bibitem [{\citenamefont {McMillan}(1968)}]{McMillan1968}%
  \BibitemOpen
  \bibfield  {author} {\bibinfo {author} {\bibfnamefont {W.~L.}\ \bibnamefont
  {McMillan}},\ }\bibfield  {title} {\bibinfo {title} {Transition temperature
  of strong-coupled superconductors},\ }\href
  {https://doi.org/10.1103/PhysRev.167.331} {\bibfield  {journal} {\bibinfo
  {journal} {Phys. Rev.}\ }\textbf {\bibinfo {volume} {167}},\ \bibinfo {pages}
  {331} (\bibinfo {year} {1968})}\BibitemShut {NoStop}%
\bibitem [{\citenamefont {Carbotte}(1990)}]{Carbotte1990}%
  \BibitemOpen
  \bibfield  {author} {\bibinfo {author} {\bibfnamefont {J.~P.}\ \bibnamefont
  {Carbotte}},\ }\bibfield  {title} {\bibinfo {title} {Properties of
  boson-exchange superconductors},\ }\href
  {https://doi.org/10.1103/RevModPhys.62.1027} {\bibfield  {journal} {\bibinfo
  {journal} {Rev. Mod. Phys.}\ }\textbf {\bibinfo {volume} {62}},\ \bibinfo
  {pages} {1027} (\bibinfo {year} {1990})}\BibitemShut {NoStop}%
\bibitem [{\citenamefont {Marsiglio}\ and\ \citenamefont
  {Carbotte}(2008)}]{Bennemann}%
  \BibitemOpen
  \bibfield  {author} {\bibinfo {author} {\bibfnamefont {F.}~\bibnamefont
  {Marsiglio}}\ and\ \bibinfo {author} {\bibfnamefont {J.~P.}\ \bibnamefont
  {Carbotte}},\ }\bibfield  {title} {\bibinfo {title} {Electron-phonon
  superconductivity},\ }in\ \href@noop {} {\emph {\bibinfo {booktitle}
  {Superconductivity, Conventional and Unconventional Superconductors}}},\
  \bibinfo {editor} {edited by\ \bibinfo {editor} {\bibfnamefont {K.~H.}\
  \bibnamefont {Bennemann}}\ and\ \bibinfo {editor} {\bibfnamefont {J.~B.}\
  \bibnamefont {Ketterson}}}\ (\bibinfo  {publisher} {Springer, Berlin},\
  \bibinfo {year} {2008})\ pp.\ \bibinfo {pages} {73--162}\BibitemShut
  {NoStop}%
\bibitem [{\citenamefont {Kresin}\ \emph {et~al.}(1984)\citenamefont {Kresin},
  \citenamefont {Gutfreund},\ and\ \citenamefont
  {Little}}]{Kresin1984Superconducting}%
  \BibitemOpen
  \bibfield  {author} {\bibinfo {author} {\bibfnamefont {V.~Z.}\ \bibnamefont
  {Kresin}}, \bibinfo {author} {\bibfnamefont {H.}~\bibnamefont {Gutfreund}},\
  and\ \bibinfo {author} {\bibfnamefont {W.~A.}\ \bibnamefont {Little}},\
  }\bibfield  {title} {\bibinfo {title} {Superconducting state in strong
  coupling},\ }\href {https://doi.org/10.1016/0038-1098(84)90701-4} {\bibfield
  {journal} {\bibinfo  {journal} {Solid State Communications}\ }\textbf
  {\bibinfo {volume} {51}},\ \bibinfo {pages} {339} (\bibinfo {year}
  {1984})}\BibitemShut {NoStop}%
\bibitem [{\citenamefont {Allen}\ and\ \citenamefont
  {Dynes}(1975)}]{Allen1975Transition}%
  \BibitemOpen
  \bibfield  {author} {\bibinfo {author} {\bibfnamefont {P.~B.}\ \bibnamefont
  {Allen}}\ and\ \bibinfo {author} {\bibfnamefont {R.~C.}\ \bibnamefont
  {Dynes}},\ }\bibfield  {title} {\bibinfo {title} {Transition temperature of
  strong-coupled superconductors reanalyzed},\ }\href
  {https://doi.org/10.1103/PhysRevB.12.905} {\bibfield  {journal} {\bibinfo
  {journal} {Phys. Rev. B}\ }\textbf {\bibinfo {volume} {12}},\ \bibinfo
  {pages} {905} (\bibinfo {year} {1975})}\BibitemShut {NoStop}%
\end{thebibliography}%


\begin{thebibliography}{68}%
\makeatletter
\providecommand \@ifxundefined [1]{%
 \@ifx{#1\undefined}
}%
\providecommand \@ifnum [1]{%
 \ifnum #1\expandafter \@firstoftwo
 \else \expandafter \@secondoftwo
 \fi
}%
\providecommand \@ifx [1]{%
 \ifx #1\expandafter \@firstoftwo
 \else \expandafter \@secondoftwo
 \fi
}%
\providecommand \natexlab [1]{#1}%
\providecommand \enquote  [1]{``#1''}%
\providecommand \bibnamefont  [1]{#1}%
\providecommand \bibfnamefont [1]{#1}%
\providecommand \citenamefont [1]{#1}%
\providecommand \href@noop [0]{\@secondoftwo}%
\providecommand \href [0]{\begingroup \@sanitize@url \@href}%
\providecommand \@href[1]{\@@startlink{#1}\@@href}%
\providecommand \@@href[1]{\endgroup#1\@@endlink}%
\providecommand \@sanitize@url [0]{\catcode `\\12\catcode `\$12\catcode
  `\&12\catcode `\#12\catcode `\^12\catcode `\_12\catcode `\%12\relax}%
\providecommand \@@startlink[1]{}%
\providecommand \@@endlink[0]{}%
\providecommand \url  [0]{\begingroup\@sanitize@url \@url }%
\providecommand \@url [1]{\endgroup\@href {#1}{\urlprefix }}%
\providecommand \urlprefix  [0]{URL }%
\providecommand \Eprint [0]{\href }%
\providecommand \doibase [0]{https://doi.org/}%
\providecommand \selectlanguage [0]{\@gobble}%
\providecommand \bibinfo  [0]{\@secondoftwo}%
\providecommand \bibfield  [0]{\@secondoftwo}%
\providecommand \translation [1]{[#1]}%
\providecommand \BibitemOpen [0]{}%
\providecommand \bibitemStop [0]{}%
\providecommand \bibitemNoStop [0]{.\EOS\space}%
\providecommand \EOS [0]{\spacefactor3000\relax}%
\providecommand \BibitemShut  [1]{\csname bibitem#1\endcsname}%
\let\auto@bib@innerbib\@empty
\bibitem [{\citenamefont {Migdal}(1958)}]{Migdal1958}%
  \BibitemOpen
  \bibfield  {author} {\bibinfo {author} {\bibfnamefont {A.~B.}\ \bibnamefont
  {Migdal}},\ }\bibfield  {title} {\bibinfo {title} {Interactions between
  electrons and lattice vibrations in a superconductor},\ }\href
  {http://jetp.ras.ru/cgi-bin/e/index/e/7/6/p996?a=list} {\bibfield  {journal}
  {\bibinfo  {journal} {Sov. Phys. JETP}\ }\textbf {\bibinfo {volume} {7}},\
  \bibinfo {pages} {996} (\bibinfo {year} {1958})}\BibitemShut {NoStop}%
\bibitem [{\citenamefont {Eliashberg}(1960)}]{Eliashberg1960}%
  \BibitemOpen
  \bibfield  {author} {\bibinfo {author} {\bibfnamefont {G.~M.}\ \bibnamefont
  {Eliashberg}},\ }\bibfield  {title} {\bibinfo {title} {Interactions between
  electrons and lattice vibrations in a superconductor},\ }\href
  {http://www.jetp.ras.ru/cgi-bin/e/index/e/11/3/p696?a=list} {\bibfield
  {journal} {\bibinfo  {journal} {Sov. Phys. JETP}\ }\textbf {\bibinfo {volume}
  {11}},\ \bibinfo {pages} {696} (\bibinfo {year} {1960})}\BibitemShut
  {NoStop}%
\bibitem [{\citenamefont {Eliashberg}(1961)}]{Eliashberg1961}%
  \BibitemOpen
  \bibfield  {author} {\bibinfo {author} {\bibfnamefont {G.~M.}\ \bibnamefont
  {Eliashberg}},\ }\bibfield  {title} {\bibinfo {title} {Temperature {G}reen's
  function for electrons in a superconductor},\ }\href
  {http://www.jetp.ras.ru/cgi-bin/e/index/e/12/5/p1000?a=list} {\bibfield
  {journal} {\bibinfo  {journal} {Sov. Phys. JETP}\ }\textbf {\bibinfo {volume}
  {12}},\ \bibinfo {pages} {1000} (\bibinfo {year} {1961})}\BibitemShut
  {NoStop}%
\bibitem [{\citenamefont {Scalapino}(1969)}]{Parks1}%
  \BibitemOpen
  \bibfield  {author} {\bibinfo {author} {\bibfnamefont {D.~J.}\ \bibnamefont
  {Scalapino}},\ }\bibfield  {title} {\bibinfo {title} {The electron-phonon
  interaction and strong-coupling superconductors},\ }in\ \href@noop {} {\emph
  {\bibinfo {booktitle} {Superconductivity: Part 1 (In Two Parts)}}},\ \bibinfo
  {editor} {edited by\ \bibinfo {editor} {\bibfnamefont {R.}~\bibnamefont
  {Parks}}}\ (\bibinfo  {publisher} {Marcel Dekker Inc., New York},\ \bibinfo
  {year} {1969})\ pp.\ \bibinfo {pages} {449--560}\BibitemShut {NoStop}%
\bibitem [{\citenamefont {Carbotte}(1990)}]{Carbotte1990}%
  \BibitemOpen
  \bibfield  {author} {\bibinfo {author} {\bibfnamefont {J.~P.}\ \bibnamefont
  {Carbotte}},\ }\bibfield  {title} {\bibinfo {title} {Properties of
  boson-exchange superconductors},\ }\href
  {https://doi.org/10.1103/RevModPhys.62.1027} {\bibfield  {journal} {\bibinfo
  {journal} {Rev. Mod. Phys.}\ }\textbf {\bibinfo {volume} {62}},\ \bibinfo
  {pages} {1027} (\bibinfo {year} {1990})}\BibitemShut {NoStop}%
\bibitem [{\citenamefont {Marsiglio}\ and\ \citenamefont
  {Carbotte}(2008)}]{Bennemann}%
  \BibitemOpen
  \bibfield  {author} {\bibinfo {author} {\bibfnamefont {F.}~\bibnamefont
  {Marsiglio}}\ and\ \bibinfo {author} {\bibfnamefont {J.~P.}\ \bibnamefont
  {Carbotte}},\ }\bibfield  {title} {\bibinfo {title} {Electron-phonon
  superconductivity},\ }in\ \href@noop {} {\emph {\bibinfo {booktitle}
  {Superconductivity, Conventional and Unconventional Superconductors}}},\
  \bibinfo {editor} {edited by\ \bibinfo {editor} {\bibfnamefont {K.~H.}\
  \bibnamefont {Bennemann}}\ and\ \bibinfo {editor} {\bibfnamefont {J.~B.}\
  \bibnamefont {Ketterson}}}\ (\bibinfo  {publisher} {Springer, Berlin},\
  \bibinfo {year} {2008})\ pp.\ \bibinfo {pages} {73--162}\BibitemShut
  {NoStop}%
\bibitem [{\citenamefont {Bardeen}\ \emph
  {et~al.}(1957{\natexlab{a}})\citenamefont {Bardeen}, \citenamefont {Cooper},\
  and\ \citenamefont {Schrieffer}}]{Bardeen1957a}%
  \BibitemOpen
  \bibfield  {author} {\bibinfo {author} {\bibfnamefont {J.}~\bibnamefont
  {Bardeen}}, \bibinfo {author} {\bibfnamefont {L.~N.}\ \bibnamefont
  {Cooper}},\ and\ \bibinfo {author} {\bibfnamefont {J.~R.}\ \bibnamefont
  {Schrieffer}},\ }\bibfield  {title} {\bibinfo {title} {Microscopic theory of
  superconductivity},\ }\href {https://doi.org/10.1103/PhysRev.106.162}
  {\bibfield  {journal} {\bibinfo  {journal} {Phys. Rev.}\ }\textbf {\bibinfo
  {volume} {106}},\ \bibinfo {pages} {162} (\bibinfo {year}
  {1957}{\natexlab{a}})}\BibitemShut {NoStop}%
\bibitem [{\citenamefont {Bardeen}\ \emph
  {et~al.}(1957{\natexlab{b}})\citenamefont {Bardeen}, \citenamefont {Cooper},\
  and\ \citenamefont {Schrieffer}}]{Bardeen1957}%
  \BibitemOpen
  \bibfield  {author} {\bibinfo {author} {\bibfnamefont {J.}~\bibnamefont
  {Bardeen}}, \bibinfo {author} {\bibfnamefont {L.~N.}\ \bibnamefont
  {Cooper}},\ and\ \bibinfo {author} {\bibfnamefont {J.~R.}\ \bibnamefont
  {Schrieffer}},\ }\bibfield  {title} {\bibinfo {title} {Theory of
  superconductivity},\ }\href {https://doi.org/10.1103/PhysRev.108.1175}
  {\bibfield  {journal} {\bibinfo  {journal} {Phys. Rev.}\ }\textbf {\bibinfo
  {volume} {108}},\ \bibinfo {pages} {1175} (\bibinfo {year}
  {1957}{\natexlab{b}})}\BibitemShut {NoStop}%
\bibitem [{\citenamefont {Bardeen}(1973)}]{Bardeen1973}%
  \BibitemOpen
  \bibfield  {author} {\bibinfo {author} {\bibfnamefont {J.}~\bibnamefont
  {Bardeen}},\ }\bibfield  {title} {\bibinfo {title} {Electron-phonon
  interactions and superconductivity},\ }\href
  {https://doi.org/10.1063/1.3128140} {\bibfield  {journal} {\bibinfo
  {journal} {Physics Today}\ }\textbf {\bibinfo {volume} {26}},\ \bibinfo
  {pages} {41} (\bibinfo {year} {1973})}\BibitemShut {NoStop}%
\bibitem [{\citenamefont {Allen}\ and\ \citenamefont
  {Dynes}(1975)}]{Allen1975Transition}%
  \BibitemOpen
  \bibfield  {author} {\bibinfo {author} {\bibfnamefont {P.~B.}\ \bibnamefont
  {Allen}}\ and\ \bibinfo {author} {\bibfnamefont {R.~C.}\ \bibnamefont
  {Dynes}},\ }\bibfield  {title} {\bibinfo {title} {Transition temperature of
  strong-coupled superconductors reanalyzed},\ }\href
  {https://doi.org/10.1103/PhysRevB.12.905} {\bibfield  {journal} {\bibinfo
  {journal} {Phys. Rev. B}\ }\textbf {\bibinfo {volume} {12}},\ \bibinfo
  {pages} {905} (\bibinfo {year} {1975})}\BibitemShut {NoStop}%
\bibitem [{\citenamefont {Allen}\ and\ \citenamefont
  {{Mitrović}}(1983)}]{Allen1983Theory}%
  \BibitemOpen
  \bibfield  {author} {\bibinfo {author} {\bibfnamefont {P.~B.}\ \bibnamefont
  {Allen}}\ and\ \bibinfo {author} {\bibfnamefont {B.}~\bibnamefont
  {{Mitrović}}},\ }\bibinfo {title} {{Theory of Superconducting $T_c$}},\ in\
  \href {https://doi.org/10.1016/s0081-1947(08)60665-7} {\emph {\bibinfo
  {booktitle} {Solid State Physics}}},\ Vol.~\bibinfo {volume} {37}\ (\bibinfo
  {publisher} {Academic Press},\ \bibinfo {year} {1983})\ p.\ \bibinfo {pages}
  {1–92}\BibitemShut {NoStop}%
\bibitem [{\citenamefont {Kresin}\ \emph {et~al.}(1984)\citenamefont {Kresin},
  \citenamefont {Gutfreund},\ and\ \citenamefont
  {Little}}]{Kresin1984Superconducting}%
  \BibitemOpen
  \bibfield  {author} {\bibinfo {author} {\bibfnamefont {V.~Z.}\ \bibnamefont
  {Kresin}}, \bibinfo {author} {\bibfnamefont {H.}~\bibnamefont {Gutfreund}},\
  and\ \bibinfo {author} {\bibfnamefont {W.~A.}\ \bibnamefont {Little}},\
  }\bibfield  {title} {\bibinfo {title} {Superconducting state in strong
  coupling},\ }\href {https://doi.org/10.1016/0038-1098(84)90701-4} {\bibfield
  {journal} {\bibinfo  {journal} {Solid State Communications}\ }\textbf
  {\bibinfo {volume} {51}},\ \bibinfo {pages} {339} (\bibinfo {year}
  {1984})}\BibitemShut {NoStop}%
\bibitem [{\citenamefont {Kresin}(1987{\natexlab{a}})}]{Kresin1987Gap}%
  \BibitemOpen
  \bibfield  {author} {\bibinfo {author} {\bibfnamefont {V.~Z.}\ \bibnamefont
  {Kresin}},\ }\bibfield  {title} {\bibinfo {title} {On the relation between
  the energy gap and the critical temperature},\ }\href
  {https://doi.org/10.1016/0038-1098(87)90119-0} {\bibfield  {journal}
  {\bibinfo  {journal} {Solid State Communications}\ }\textbf {\bibinfo
  {volume} {63}},\ \bibinfo {pages} {725} (\bibinfo {year}
  {1987}{\natexlab{a}})}\BibitemShut {NoStop}%
\bibitem [{\citenamefont {Kresin}(1987{\natexlab{b}})}]{Kresin1987Tc}%
  \BibitemOpen
  \bibfield  {author} {\bibinfo {author} {\bibfnamefont {V.~Z.}\ \bibnamefont
  {Kresin}},\ }\bibfield  {title} {\bibinfo {title} {On the critical
  temperature for any strength of the electron-phonon coupling},\ }\href
  {https://doi.org/10.1016/0375-9601(87)90744-4} {\bibfield  {journal}
  {\bibinfo  {journal} {Physics Letters A}\ }\textbf {\bibinfo {volume}
  {122}},\ \bibinfo {pages} {434} (\bibinfo {year}
  {1987}{\natexlab{b}})}\BibitemShut {NoStop}%
\bibitem [{\citenamefont
  {Combescot}(1989{\natexlab{a}})}]{Combescot1989Spectral}%
  \BibitemOpen
  \bibfield  {author} {\bibinfo {author} {\bibfnamefont {R.}~\bibnamefont
  {Combescot}},\ }\bibfield  {title} {\bibinfo {title} {On the spectral
  dependence of the critical temperature of superconductors},\ }\href
  {https://doi.org/10.1016/0921-4534(89)90795-8} {\bibfield  {journal}
  {\bibinfo  {journal} {Physica C: Superconductivity and its Applications}\
  }\textbf {\bibinfo {volume} {162}},\ \bibinfo {pages} {1507} (\bibinfo {year}
  {1989}{\natexlab{a}})}\BibitemShut {NoStop}%
\bibitem [{\citenamefont {Combescot}(1989{\natexlab{b}})}]{Combescot1989Tc}%
  \BibitemOpen
  \bibfield  {author} {\bibinfo {author} {\bibfnamefont {R.}~\bibnamefont
  {Combescot}},\ }\bibfield  {title} {\bibinfo {title} {Critical temperature of
  superconductors: The spectral dependence},\ }\href
  {https://doi.org/10.1209/0295-5075/10/2/015} {\bibfield  {journal} {\bibinfo
  {journal} {Europhysics Letters (EPL)}\ }\textbf {\bibinfo {volume} {10}},\
  \bibinfo {pages} {177} (\bibinfo {year} {1989}{\natexlab{b}})}\BibitemShut
  {NoStop}%
\bibitem [{\citenamefont {Marsiglio}(2020)}]{Marsiglio2020Eliashberg}%
  \BibitemOpen
  \bibfield  {author} {\bibinfo {author} {\bibfnamefont {F.}~\bibnamefont
  {Marsiglio}},\ }\bibfield  {title} {\bibinfo {title} {{Eliashberg theory: A
  short review}},\ }\href {https://doi.org/10.1016/j.aop.2020.168102}
  {\bibfield  {journal} {\bibinfo  {journal} {Ann. Phys.}\ }\textbf {\bibinfo
  {volume} {417}},\ \bibinfo {pages} {168102} (\bibinfo {year}
  {2020})}\BibitemShut {NoStop}%
\bibitem [{\citenamefont {Protter}\ \emph {et~al.}(2021)\citenamefont
  {Protter}, \citenamefont {Boyack},\ and\ \citenamefont
  {Marsiglio}}]{Protter2021}%
  \BibitemOpen
  \bibfield  {author} {\bibinfo {author} {\bibfnamefont {M.}~\bibnamefont
  {Protter}}, \bibinfo {author} {\bibfnamefont {R.}~\bibnamefont {Boyack}},\
  and\ \bibinfo {author} {\bibfnamefont {F.}~\bibnamefont {Marsiglio}},\
  }\bibfield  {title} {\bibinfo {title} {Functional-integral approach to
  {G}aussian fluctuations in {E}liashberg theory},\ }\href
  {https://doi.org/10.1103/PhysRevB.104.014513} {\bibfield  {journal} {\bibinfo
   {journal} {Phys. Rev. B}\ }\textbf {\bibinfo {volume} {104}},\ \bibinfo
  {pages} {014513} (\bibinfo {year} {2021})}\BibitemShut {NoStop}%
\bibitem [{\citenamefont {Alexandrov}(2001)}]{Alexandrov2001Breakdown}%
  \BibitemOpen
  \bibfield  {author} {\bibinfo {author} {\bibfnamefont {A.~S.}\ \bibnamefont
  {Alexandrov}},\ }\bibfield  {title} {\bibinfo {title} {Breakdown of the
  {M}igdal-{E}liashberg theory in the strong-coupling adiabatic regime},\
  }\href {https://doi.org/10.1209/epl/i2001-00492-x} {\bibfield  {journal}
  {\bibinfo  {journal} {EPL}\ }\textbf {\bibinfo {volume} {56}},\ \bibinfo
  {pages} {92} (\bibinfo {year} {2001})}\BibitemShut {NoStop}%
\bibitem [{\citenamefont {Esterlis}\ \emph {et~al.}(2018)\citenamefont
  {Esterlis}, \citenamefont {Nosarzewski}, \citenamefont {Huang}, \citenamefont
  {Moritz}, \citenamefont {Devereaux}, \citenamefont {Scalapino},\ and\
  \citenamefont {Kivelson}}]{Esterlis2018Breakdown}%
  \BibitemOpen
  \bibfield  {author} {\bibinfo {author} {\bibfnamefont {I.}~\bibnamefont
  {Esterlis}}, \bibinfo {author} {\bibfnamefont {B.}~\bibnamefont
  {Nosarzewski}}, \bibinfo {author} {\bibfnamefont {E.~W.}\ \bibnamefont
  {Huang}}, \bibinfo {author} {\bibfnamefont {B.}~\bibnamefont {Moritz}},
  \bibinfo {author} {\bibfnamefont {T.~P.}\ \bibnamefont {Devereaux}}, \bibinfo
  {author} {\bibfnamefont {D.~J.}\ \bibnamefont {Scalapino}},\ and\ \bibinfo
  {author} {\bibfnamefont {S.~A.}\ \bibnamefont {Kivelson}},\ }\bibfield
  {title} {\bibinfo {title} {Breakdown of the {M}igdal-{E}liashberg theory: A
  determinant quantum {M}onte {C}arlo study},\ }\href
  {https://doi.org/10.1103/PhysRevB.97.140501} {\bibfield  {journal} {\bibinfo
  {journal} {Phys. Rev. B}\ }\textbf {\bibinfo {volume} {97}},\ \bibinfo
  {pages} {140501} (\bibinfo {year} {2018})}\BibitemShut {NoStop}%
\bibitem [{\citenamefont {Chubukov}\ \emph {et~al.}(2020)\citenamefont
  {Chubukov}, \citenamefont {Abanov}, \citenamefont {Esterlis},\ and\
  \citenamefont {Kivelson}}]{Chubukov2020Eliashberg}%
  \BibitemOpen
  \bibfield  {author} {\bibinfo {author} {\bibfnamefont {A.~V.}\ \bibnamefont
  {Chubukov}}, \bibinfo {author} {\bibfnamefont {A.}~\bibnamefont {Abanov}},
  \bibinfo {author} {\bibfnamefont {I.}~\bibnamefont {Esterlis}},\ and\
  \bibinfo {author} {\bibfnamefont {S.~A.}\ \bibnamefont {Kivelson}},\
  }\bibfield  {title} {\bibinfo {title} {Eliashberg theory of phonon-mediated
  superconductivity — when it is valid and how it breaks down},\ }\href
  {https://doi.org/10.1016/j.aop.2020.168190} {\bibfield  {journal} {\bibinfo
  {journal} {Annals of Physics}\ }\textbf {\bibinfo {volume} {417}},\ \bibinfo
  {pages} {168190} (\bibinfo {year} {2020})}\BibitemShut {NoStop}%
\bibitem [{\citenamefont {Yuzbashyan}\ and\ \citenamefont
  {Altshuler}(2022)}]{Yuzbashyan2022}%
  \BibitemOpen
  \bibfield  {author} {\bibinfo {author} {\bibfnamefont {E.~A.}\ \bibnamefont
  {Yuzbashyan}}\ and\ \bibinfo {author} {\bibfnamefont {B.~L.}\ \bibnamefont
  {Altshuler}},\ }\bibfield  {title} {\bibinfo {title} {Breakdown of the
  {M}igdal-{E}liashberg theory and a theory of lattice-fermionic
  superfluidity},\ }\href {https://doi.org/10.1103/PhysRevB.106.054518}
  {\bibfield  {journal} {\bibinfo  {journal} {Phys. Rev. B}\ }\textbf {\bibinfo
  {volume} {106}},\ \bibinfo {pages} {054518} (\bibinfo {year}
  {2022})}\BibitemShut {NoStop}%
\bibitem [{\citenamefont {Zhang}\ \emph {et~al.}(2024)\citenamefont {Zhang},
  \citenamefont {Raines},\ and\ \citenamefont
  {Chubukov}}]{Zhang2024Applicability}%
  \BibitemOpen
  \bibfield  {author} {\bibinfo {author} {\bibfnamefont {S.-S.}\ \bibnamefont
  {Zhang}}, \bibinfo {author} {\bibfnamefont {Z.~M.}\ \bibnamefont {Raines}},\
  and\ \bibinfo {author} {\bibfnamefont {A.~V.}\ \bibnamefont {Chubukov}},\
  }\bibfield  {title} {\bibinfo {title} {Applicability of {E}liashberg theory
  for systems with electron-phonon and electron-electron interaction: A
  comparative analysis},\ }\href {https://doi.org/10.1103/PhysRevB.109.245132}
  {\bibfield  {journal} {\bibinfo  {journal} {Phys. Rev. B}\ }\textbf {\bibinfo
  {volume} {109}},\ \bibinfo {pages} {245132} (\bibinfo {year}
  {2024})}\BibitemShut {NoStop}%
\bibitem [{\citenamefont {Millis}(1992)}]{Millis1992Nearly}%
  \BibitemOpen
  \bibfield  {author} {\bibinfo {author} {\bibfnamefont {A.~J.}\ \bibnamefont
  {Millis}},\ }\bibfield  {title} {\bibinfo {title} {Nearly antiferromagnetic
  {F}ermi liquids: An analytic {E}liashberg approach},\ }\href
  {https://doi.org/10.1103/PhysRevB.45.13047} {\bibfield  {journal} {\bibinfo
  {journal} {Phys. Rev. B}\ }\textbf {\bibinfo {volume} {45}},\ \bibinfo
  {pages} {13047} (\bibinfo {year} {1992})}\BibitemShut {NoStop}%
\bibitem [{\citenamefont {Rech}\ \emph {et~al.}(2006)\citenamefont {Rech},
  \citenamefont {P\'epin},\ and\ \citenamefont {Chubukov}}]{Rech2006Quantum}%
  \BibitemOpen
  \bibfield  {author} {\bibinfo {author} {\bibfnamefont {J.}~\bibnamefont
  {Rech}}, \bibinfo {author} {\bibfnamefont {C.}~\bibnamefont {P\'epin}},\ and\
  \bibinfo {author} {\bibfnamefont {A.~V.}\ \bibnamefont {Chubukov}},\
  }\bibfield  {title} {\bibinfo {title} {Quantum critical behavior in itinerant
  electron systems: Eliashberg theory and instability of a ferromagnetic
  quantum critical point},\ }\href {https://doi.org/10.1103/PhysRevB.74.195126}
  {\bibfield  {journal} {\bibinfo  {journal} {Phys. Rev. B}\ }\textbf {\bibinfo
  {volume} {74}},\ \bibinfo {pages} {195126} (\bibinfo {year}
  {2006})}\BibitemShut {NoStop}%
\bibitem [{\citenamefont {Phillips}\ \emph {et~al.}(2022)\citenamefont
  {Phillips}, \citenamefont {Hussey},\ and\ \citenamefont
  {Abbamonte}}]{Phillips2022}%
  \BibitemOpen
  \bibfield  {author} {\bibinfo {author} {\bibfnamefont {P.~W.}\ \bibnamefont
  {Phillips}}, \bibinfo {author} {\bibfnamefont {N.~E.}\ \bibnamefont
  {Hussey}},\ and\ \bibinfo {author} {\bibfnamefont {P.}~\bibnamefont
  {Abbamonte}},\ }\bibfield  {title} {\bibinfo {title} {{Stranger than
  metals}},\ }\href {https://doi.org/10.1126/science.abh4273} {\bibfield
  {journal} {\bibinfo  {journal} {Science}\ }\textbf {\bibinfo {volume}
  {377}},\ \bibinfo {pages} {6602} (\bibinfo {year} {2022})}\BibitemShut
  {NoStop}%
\bibitem [{\citenamefont {Patel}\ \emph {et~al.}(2023)\citenamefont {Patel},
  \citenamefont {Guo}, \citenamefont {Esterlis},\ and\ \citenamefont
  {Sachdev}}]{Patel2023Universal}%
  \BibitemOpen
  \bibfield  {author} {\bibinfo {author} {\bibfnamefont {A.~A.}\ \bibnamefont
  {Patel}}, \bibinfo {author} {\bibfnamefont {H.}~\bibnamefont {Guo}}, \bibinfo
  {author} {\bibfnamefont {I.}~\bibnamefont {Esterlis}},\ and\ \bibinfo
  {author} {\bibfnamefont {S.}~\bibnamefont {Sachdev}},\ }\bibfield  {title}
  {\bibinfo {title} {Universal theory of strange metals from spatially random
  interactions},\ }\href {https://doi.org/10.1126/science.abq6011} {\bibfield
  {journal} {\bibinfo  {journal} {Science}\ }\textbf {\bibinfo {volume}
  {381}},\ \bibinfo {pages} {790} (\bibinfo {year} {2023})}\BibitemShut
  {NoStop}%
\bibitem [{\citenamefont {Li}\ \emph {et~al.}(2024)\citenamefont {Li},
  \citenamefont {Valentinis}, \citenamefont {Patel}, \citenamefont {Guo},
  \citenamefont {Schmalian}, \citenamefont {Sachdev},\ and\ \citenamefont
  {Esterlis}}]{Li2024Strange}%
  \BibitemOpen
  \bibfield  {author} {\bibinfo {author} {\bibfnamefont {C.}~\bibnamefont
  {Li}}, \bibinfo {author} {\bibfnamefont {D.}~\bibnamefont {Valentinis}},
  \bibinfo {author} {\bibfnamefont {A.~A.}\ \bibnamefont {Patel}}, \bibinfo
  {author} {\bibfnamefont {H.}~\bibnamefont {Guo}}, \bibinfo {author}
  {\bibfnamefont {J.}~\bibnamefont {Schmalian}}, \bibinfo {author}
  {\bibfnamefont {S.}~\bibnamefont {Sachdev}},\ and\ \bibinfo {author}
  {\bibfnamefont {I.}~\bibnamefont {Esterlis}},\ }\bibfield  {title} {\bibinfo
  {title} {Strange metal and superconductor in the two-dimensional
  {Yukawa-Sachdev-Ye-Kitaev} model},\ }\href
  {https://doi.org/10.1103/PhysRevLett.133.186502} {\bibfield  {journal}
  {\bibinfo  {journal} {Phys. Rev. Lett.}\ }\textbf {\bibinfo {volume} {133}},\
  \bibinfo {pages} {186502} (\bibinfo {year} {2024})}\BibitemShut {NoStop}%
\bibitem [{\citenamefont {Esterlis}\ and\ \citenamefont
  {Schmalian}(2019)}]{Esterlis2019Cooper}%
  \BibitemOpen
  \bibfield  {author} {\bibinfo {author} {\bibfnamefont {I.}~\bibnamefont
  {Esterlis}}\ and\ \bibinfo {author} {\bibfnamefont {J.}~\bibnamefont
  {Schmalian}},\ }\bibfield  {title} {\bibinfo {title} {Cooper pairing of
  incoherent electrons: An electron-phonon version of the {Sachdev-Ye-Kitaev}
  model},\ }\href {https://doi.org/10.1103/PhysRevB.100.115132} {\bibfield
  {journal} {\bibinfo  {journal} {Phys. Rev. B}\ }\textbf {\bibinfo {volume}
  {100}},\ \bibinfo {pages} {115132} (\bibinfo {year} {2019})}\BibitemShut
  {NoStop}%
\bibitem [{\citenamefont {Wang}(2020)}]{Wang2020Solvable}%
  \BibitemOpen
  \bibfield  {author} {\bibinfo {author} {\bibfnamefont {Y.}~\bibnamefont
  {Wang}},\ }\bibfield  {title} {\bibinfo {title} {Solvable strong-coupling
  quantum-dot model with a non-{F}ermi-liquid pairing transition},\ }\href
  {https://doi.org/10.1103/PhysRevLett.124.017002} {\bibfield  {journal}
  {\bibinfo  {journal} {Phys. Rev. Lett.}\ }\textbf {\bibinfo {volume} {124}},\
  \bibinfo {pages} {017002} (\bibinfo {year} {2020})}\BibitemShut {NoStop}%
\bibitem [{\citenamefont {Wang}\ and\ \citenamefont
  {Chubukov}(2020)}]{Wang2020Quantum}%
  \BibitemOpen
  \bibfield  {author} {\bibinfo {author} {\bibfnamefont {Y.}~\bibnamefont
  {Wang}}\ and\ \bibinfo {author} {\bibfnamefont {A.~V.}\ \bibnamefont
  {Chubukov}},\ }\bibfield  {title} {\bibinfo {title} {Quantum phase transition
  in the {Yukawa-SYK} model},\ }\href
  {https://doi.org/10.1103/PhysRevResearch.2.033084} {\bibfield  {journal}
  {\bibinfo  {journal} {Phys. Rev. Res.}\ }\textbf {\bibinfo {volume} {2}},\
  \bibinfo {pages} {033084} (\bibinfo {year} {2020})}\BibitemShut {NoStop}%
\bibitem [{\citenamefont {Wang}\ \emph {et~al.}(2021)\citenamefont {Wang},
  \citenamefont {Davis}, \citenamefont {Pan}, \citenamefont {Wang},\ and\
  \citenamefont {Meng}}]{Wang2021Phase}%
  \BibitemOpen
  \bibfield  {author} {\bibinfo {author} {\bibfnamefont {W.}~\bibnamefont
  {Wang}}, \bibinfo {author} {\bibfnamefont {A.}~\bibnamefont {Davis}},
  \bibinfo {author} {\bibfnamefont {G.}~\bibnamefont {Pan}}, \bibinfo {author}
  {\bibfnamefont {Y.}~\bibnamefont {Wang}},\ and\ \bibinfo {author}
  {\bibfnamefont {Z.~Y.}\ \bibnamefont {Meng}},\ }\bibfield  {title} {\bibinfo
  {title} {Phase diagram of the spin-$\frac{1}{2}$ {Yukawa--Sachdev-Ye-Kitaev}
  model: Non-{F}ermi liquid, insulator, and superconductor},\ }\href
  {https://doi.org/10.1103/PhysRevB.103.195108} {\bibfield  {journal} {\bibinfo
   {journal} {Phys. Rev. B}\ }\textbf {\bibinfo {volume} {103}},\ \bibinfo
  {pages} {195108} (\bibinfo {year} {2021})}\BibitemShut {NoStop}%
\bibitem [{\citenamefont {Pan}\ \emph {et~al.}(2021)\citenamefont {Pan},
  \citenamefont {Wang}, \citenamefont {Davis}, \citenamefont {Wang},\ and\
  \citenamefont {Meng}}]{Pan2021Yukawa}%
  \BibitemOpen
  \bibfield  {author} {\bibinfo {author} {\bibfnamefont {G.}~\bibnamefont
  {Pan}}, \bibinfo {author} {\bibfnamefont {W.}~\bibnamefont {Wang}}, \bibinfo
  {author} {\bibfnamefont {A.}~\bibnamefont {Davis}}, \bibinfo {author}
  {\bibfnamefont {Y.}~\bibnamefont {Wang}},\ and\ \bibinfo {author}
  {\bibfnamefont {Z.~Y.}\ \bibnamefont {Meng}},\ }\bibfield  {title} {\bibinfo
  {title} {{Yukawa-SYK} model and self-tuned quantum criticality},\ }\href
  {https://doi.org/10.1103/PhysRevResearch.3.013250} {\bibfield  {journal}
  {\bibinfo  {journal} {Phys. Rev. Res.}\ }\textbf {\bibinfo {volume} {3}},\
  \bibinfo {pages} {013250} (\bibinfo {year} {2021})}\BibitemShut {NoStop}%
\bibitem [{\citenamefont {Esterlis}\ \emph {et~al.}(2021)\citenamefont
  {Esterlis}, \citenamefont {Guo}, \citenamefont {Patel},\ and\ \citenamefont
  {Sachdev}}]{Esterlis2021Large}%
  \BibitemOpen
  \bibfield  {author} {\bibinfo {author} {\bibfnamefont {I.}~\bibnamefont
  {Esterlis}}, \bibinfo {author} {\bibfnamefont {H.}~\bibnamefont {Guo}},
  \bibinfo {author} {\bibfnamefont {A.~A.}\ \bibnamefont {Patel}},\ and\
  \bibinfo {author} {\bibfnamefont {S.}~\bibnamefont {Sachdev}},\ }\bibfield
  {title} {\bibinfo {title} {Large-{$N$} theory of critical {F}ermi surfaces},\
  }\href {https://doi.org/10.1103/PhysRevB.103.235129} {\bibfield  {journal}
  {\bibinfo  {journal} {Phys. Rev. B}\ }\textbf {\bibinfo {volume} {103}},\
  \bibinfo {pages} {235129} (\bibinfo {year} {2021})}\BibitemShut {NoStop}%
\bibitem [{\citenamefont {Guo}\ \emph {et~al.}(2022)\citenamefont {Guo},
  \citenamefont {Patel}, \citenamefont {Esterlis},\ and\ \citenamefont
  {Sachdev}}]{Guo2022Large}%
  \BibitemOpen
  \bibfield  {author} {\bibinfo {author} {\bibfnamefont {H.}~\bibnamefont
  {Guo}}, \bibinfo {author} {\bibfnamefont {A.~A.}\ \bibnamefont {Patel}},
  \bibinfo {author} {\bibfnamefont {I.}~\bibnamefont {Esterlis}},\ and\
  \bibinfo {author} {\bibfnamefont {S.}~\bibnamefont {Sachdev}},\ }\bibfield
  {title} {\bibinfo {title} {Large-{$N$} theory of critical {Fermi} surfaces.
  {II}. {C}onductivity},\ }\href {https://doi.org/10.1103/PhysRevB.106.115151}
  {\bibfield  {journal} {\bibinfo  {journal} {Phys. Rev. B}\ }\textbf {\bibinfo
  {volume} {106}},\ \bibinfo {pages} {115151} (\bibinfo {year}
  {2022})}\BibitemShut {NoStop}%
\bibitem [{\citenamefont {Guo}\ \emph {et~al.}(2024)\citenamefont {Guo},
  \citenamefont {Valentinis}, \citenamefont {Schmalian}, \citenamefont
  {Sachdev},\ and\ \citenamefont {Patel}}]{Guo2024Cyclotron}%
  \BibitemOpen
  \bibfield  {author} {\bibinfo {author} {\bibfnamefont {H.}~\bibnamefont
  {Guo}}, \bibinfo {author} {\bibfnamefont {D.}~\bibnamefont {Valentinis}},
  \bibinfo {author} {\bibfnamefont {J.}~\bibnamefont {Schmalian}}, \bibinfo
  {author} {\bibfnamefont {S.}~\bibnamefont {Sachdev}},\ and\ \bibinfo {author}
  {\bibfnamefont {A.~A.}\ \bibnamefont {Patel}},\ }\bibfield  {title} {\bibinfo
  {title} {Cyclotron resonance and quantum oscillations of critical {F}ermi
  surfaces},\ }\href {https://doi.org/10.1103/PhysRevB.109.075162} {\bibfield
  {journal} {\bibinfo  {journal} {Phys. Rev. B}\ }\textbf {\bibinfo {volume}
  {109}},\ \bibinfo {pages} {075162} (\bibinfo {year} {2024})}\BibitemShut
  {NoStop}%
\bibitem [{\citenamefont {Wang}\ \emph {et~al.}(2025)\citenamefont {Wang},
  \citenamefont {Ge},\ and\ \citenamefont {Sin}}]{Wang2025Linear}%
  \BibitemOpen
  \bibfield  {author} {\bibinfo {author} {\bibfnamefont {Y.-L.}\ \bibnamefont
  {Wang}}, \bibinfo {author} {\bibfnamefont {X.-H.}\ \bibnamefont {Ge}},\ and\
  \bibinfo {author} {\bibfnamefont {S.-J.}\ \bibnamefont {Sin}},\ }\bibfield
  {title} {\bibinfo {title} {Linear-{$T$} resistivity from spatially random
  vector coupling},\ }\href {https://doi.org/10.1103/PhysRevB.111.115135}
  {\bibfield  {journal} {\bibinfo  {journal} {Phys. Rev. B}\ }\textbf {\bibinfo
  {volume} {111}},\ \bibinfo {pages} {115135} (\bibinfo {year}
  {2025})}\BibitemShut {NoStop}%
\bibitem [{\citenamefont {Hauck}\ \emph {et~al.}(2020)\citenamefont {Hauck},
  \citenamefont {Klug}, \citenamefont {Esterlis},\ and\ \citenamefont
  {Schmalian}}]{Hauck2020Eliashberg}%
  \BibitemOpen
  \bibfield  {author} {\bibinfo {author} {\bibfnamefont {D.}~\bibnamefont
  {Hauck}}, \bibinfo {author} {\bibfnamefont {M.~J.}\ \bibnamefont {Klug}},
  \bibinfo {author} {\bibfnamefont {I.}~\bibnamefont {Esterlis}},\ and\
  \bibinfo {author} {\bibfnamefont {J.}~\bibnamefont {Schmalian}},\ }\bibfield
  {title} {\bibinfo {title} {Eliashberg equations for an electron-phonon
  version of the {Sachdev-Ye-Kitaev} model: Pair breaking in non-{F}ermi liquid
  superconductors},\ }\href
  {https://doi.org/https://doi.org/10.1016/j.aop.2020.168120} {\bibfield
  {journal} {\bibinfo  {journal} {Annals of Physics}\ }\textbf {\bibinfo
  {volume} {417}},\ \bibinfo {pages} {168120} (\bibinfo {year}
  {2020})}\BibitemShut {NoStop}%
\bibitem [{\citenamefont {Classen}\ and\ \citenamefont
  {Chubukov}(2021)}]{Classen2021Superconductivity}%
  \BibitemOpen
  \bibfield  {author} {\bibinfo {author} {\bibfnamefont {L.}~\bibnamefont
  {Classen}}\ and\ \bibinfo {author} {\bibfnamefont {A.}~\bibnamefont
  {Chubukov}},\ }\bibfield  {title} {\bibinfo {title} {Superconductivity of
  incoherent electrons in the {Yukawa Sachdev-Ye-Kitaev} model},\ }\href
  {https://doi.org/10.1103/PhysRevB.104.125120} {\bibfield  {journal} {\bibinfo
   {journal} {Phys. Rev. B}\ }\textbf {\bibinfo {volume} {104}},\ \bibinfo
  {pages} {125120} (\bibinfo {year} {2021})}\BibitemShut {NoStop}%
\bibitem [{\citenamefont {Choi}\ \emph {et~al.}(2022)\citenamefont {Choi},
  \citenamefont {Tavakol},\ and\ \citenamefont {Kim}}]{Choi2022Pairing}%
  \BibitemOpen
  \bibfield  {author} {\bibinfo {author} {\bibfnamefont {W.}~\bibnamefont
  {Choi}}, \bibinfo {author} {\bibfnamefont {O.}~\bibnamefont {Tavakol}},\ and\
  \bibinfo {author} {\bibfnamefont {Y.~B.}\ \bibnamefont {Kim}},\ }\bibfield
  {title} {\bibinfo {title} {Pairing instabilities of the {Yukawa-SYK} models
  with controlled fermion incoherence},\ }\href
  {https://doi.org/10.21468/SciPostPhys.12.5.151} {\bibfield  {journal}
  {\bibinfo  {journal} {SciPost Phys.}\ }\textbf {\bibinfo {volume} {12}},\
  \bibinfo {pages} {151} (\bibinfo {year} {2022})}\BibitemShut {NoStop}%
\bibitem [{\citenamefont {Valentinis}\ \emph
  {et~al.}(2023{\natexlab{a}})\citenamefont {Valentinis}, \citenamefont
  {Inkof},\ and\ \citenamefont {Schmalian}}]{Valentinis2023BCS}%
  \BibitemOpen
  \bibfield  {author} {\bibinfo {author} {\bibfnamefont {D.}~\bibnamefont
  {Valentinis}}, \bibinfo {author} {\bibfnamefont {G.~A.}\ \bibnamefont
  {Inkof}},\ and\ \bibinfo {author} {\bibfnamefont {J.}~\bibnamefont
  {Schmalian}},\ }\bibfield  {title} {\bibinfo {title} {{BCS} to incoherent
  superconductivity crossover in the {Yukawa-Sachdev-Ye-Kitaev} model on a
  lattice},\ }\href {https://doi.org/10.1103/PhysRevB.108.L140501} {\bibfield
  {journal} {\bibinfo  {journal} {Phys. Rev. B}\ }\textbf {\bibinfo {volume}
  {108}},\ \bibinfo {pages} {L140501} (\bibinfo {year}
  {2023}{\natexlab{a}})}\BibitemShut {NoStop}%
\bibitem [{\citenamefont {Valentinis}\ \emph
  {et~al.}(2023{\natexlab{b}})\citenamefont {Valentinis}, \citenamefont
  {Inkof},\ and\ \citenamefont {Schmalian}}]{Valentinis2023Correlation}%
  \BibitemOpen
  \bibfield  {author} {\bibinfo {author} {\bibfnamefont {D.}~\bibnamefont
  {Valentinis}}, \bibinfo {author} {\bibfnamefont {G.~A.}\ \bibnamefont
  {Inkof}},\ and\ \bibinfo {author} {\bibfnamefont {J.}~\bibnamefont
  {Schmalian}},\ }\bibfield  {title} {\bibinfo {title} {Correlation between
  phase stiffness and condensation energy across the non-{F}ermi to
  {F}ermi-liquid crossover in the {Yukawa-Sachdev-Ye-Kitaev} model on a
  lattice},\ }\href {https://doi.org/10.1103/PhysRevResearch.5.043007}
  {\bibfield  {journal} {\bibinfo  {journal} {Phys. Rev. Res.}\ }\textbf
  {\bibinfo {volume} {5}},\ \bibinfo {pages} {043007} (\bibinfo {year}
  {2023}{\natexlab{b}})}\BibitemShut {NoStop}%
\bibitem [{\citenamefont {Esterlis}\ and\ \citenamefont
  {Schmalian}(2025)}]{Esterlis2025QCETh}%
  \BibitemOpen
  \bibfield  {author} {\bibinfo {author} {\bibfnamefont {I.}~\bibnamefont
  {Esterlis}}\ and\ \bibinfo {author} {\bibfnamefont {J.}~\bibnamefont
  {Schmalian}},\ }\href {https://arxiv.org/abs/2506.11952} {\bibinfo {title}
  {Quantum critical {E}liashberg theory}} (\bibinfo {year} {2025}),\ \Eprint
  {https://arxiv.org/abs/2506.11952} {arXiv:2506.11952 [cond-mat.str-el]}
  \BibitemShut {NoStop}%
\bibitem [{\citenamefont {Hazra}\ \emph {et~al.}(2019)\citenamefont {Hazra},
  \citenamefont {Verma},\ and\ \citenamefont {Randeria}}]{Hazra2019Bounds}%
  \BibitemOpen
  \bibfield  {author} {\bibinfo {author} {\bibfnamefont {T.}~\bibnamefont
  {Hazra}}, \bibinfo {author} {\bibfnamefont {N.}~\bibnamefont {Verma}},\ and\
  \bibinfo {author} {\bibfnamefont {M.}~\bibnamefont {Randeria}},\ }\bibfield
  {title} {\bibinfo {title} {Bounds on the superconducting transition
  temperature: Applications to twisted bilayer graphene and cold atoms},\
  }\href {https://doi.org/10.1103/PhysRevX.9.031049} {\bibfield  {journal}
  {\bibinfo  {journal} {Phys. Rev. X}\ }\textbf {\bibinfo {volume} {9}},\
  \bibinfo {pages} {031049} (\bibinfo {year} {2019})}\BibitemShut {NoStop}%
\bibitem [{\citenamefont {Abanov}\ and\ \citenamefont
  {Chubukov}(2020)}]{Abanov2020gamma01}%
  \BibitemOpen
  \bibfield  {author} {\bibinfo {author} {\bibfnamefont {A.}~\bibnamefont
  {Abanov}}\ and\ \bibinfo {author} {\bibfnamefont {A.~V.}\ \bibnamefont
  {Chubukov}},\ }\bibfield  {title} {\bibinfo {title} {Interplay between
  superconductivity and non-{F}ermi liquid at a quantum critical point in a
  metal. {I}. {T}he {$\ensuremath{\gamma}$} model and its phase diagram at
  {$T=0$}: The case {$0<\ensuremath{\gamma}<1$}},\ }\href
  {https://doi.org/10.1103/PhysRevB.102.024524} {\bibfield  {journal} {\bibinfo
   {journal} {Phys. Rev. B}\ }\textbf {\bibinfo {volume} {102}},\ \bibinfo
  {pages} {024524} (\bibinfo {year} {2020})}\BibitemShut {NoStop}%
\bibitem [{\citenamefont {Wu}\ \emph {et~al.}(2020{\natexlab{a}})\citenamefont
  {Wu}, \citenamefont {Abanov}, \citenamefont {Wang},\ and\ \citenamefont
  {Chubukov}}]{Wu2020gamma01T}%
  \BibitemOpen
  \bibfield  {author} {\bibinfo {author} {\bibfnamefont {Y.-M.}\ \bibnamefont
  {Wu}}, \bibinfo {author} {\bibfnamefont {A.}~\bibnamefont {Abanov}}, \bibinfo
  {author} {\bibfnamefont {Y.}~\bibnamefont {Wang}},\ and\ \bibinfo {author}
  {\bibfnamefont {A.~V.}\ \bibnamefont {Chubukov}},\ }\bibfield  {title}
  {\bibinfo {title} {Interplay between superconductivity and non-{F}ermi liquid
  at a quantum critical point in a metal. {II}. {T}he {$\ensuremath{\gamma}$}
  model at a finite {$T$} for {$0<\ensuremath{\gamma}<1$}},\ }\href
  {https://doi.org/10.1103/PhysRevB.102.024525} {\bibfield  {journal} {\bibinfo
   {journal} {Phys. Rev. B}\ }\textbf {\bibinfo {volume} {102}},\ \bibinfo
  {pages} {024525} (\bibinfo {year} {2020}{\natexlab{a}})}\BibitemShut
  {NoStop}%
\bibitem [{\citenamefont {Wu}\ \emph {et~al.}(2020{\natexlab{b}})\citenamefont
  {Wu}, \citenamefont {Abanov},\ and\ \citenamefont {Chubukov}}]{Wu2020gamma1}%
  \BibitemOpen
  \bibfield  {author} {\bibinfo {author} {\bibfnamefont {Y.-M.}\ \bibnamefont
  {Wu}}, \bibinfo {author} {\bibfnamefont {A.}~\bibnamefont {Abanov}},\ and\
  \bibinfo {author} {\bibfnamefont {A.~V.}\ \bibnamefont {Chubukov}},\
  }\bibfield  {title} {\bibinfo {title} {Interplay between superconductivity
  and non-{F}ermi liquid behavior at a quantum critical point in a metal.
  {III}. {T}he $\ensuremath{\gamma}$ model and its phase diagram across
  $\ensuremath{\gamma}=1$},\ }\href
  {https://doi.org/10.1103/PhysRevB.102.094516} {\bibfield  {journal} {\bibinfo
   {journal} {Phys. Rev. B}\ }\textbf {\bibinfo {volume} {102}},\ \bibinfo
  {pages} {094516} (\bibinfo {year} {2020}{\natexlab{b}})}\BibitemShut
  {NoStop}%
\bibitem [{\citenamefont {Wu}\ \emph {et~al.}(2021)\citenamefont {Wu},
  \citenamefont {Zhang}, \citenamefont {Abanov},\ and\ \citenamefont
  {Chubukov}}]{Wu2021gamma2}%
  \BibitemOpen
  \bibfield  {author} {\bibinfo {author} {\bibfnamefont {Y.-M.}\ \bibnamefont
  {Wu}}, \bibinfo {author} {\bibfnamefont {S.-S.}\ \bibnamefont {Zhang}},
  \bibinfo {author} {\bibfnamefont {A.}~\bibnamefont {Abanov}},\ and\ \bibinfo
  {author} {\bibfnamefont {A.~V.}\ \bibnamefont {Chubukov}},\ }\bibfield
  {title} {\bibinfo {title} {Interplay between superconductivity and
  non-{F}ermi liquid behavior at a quantum-critical point in a metal. {V.}
  {T}he $\ensuremath{\gamma}$ model and its phase diagram: The case
  $\ensuremath{\gamma}=2$},\ }\href
  {https://doi.org/10.1103/PhysRevB.103.184508} {\bibfield  {journal} {\bibinfo
   {journal} {Phys. Rev. B}\ }\textbf {\bibinfo {volume} {103}},\ \bibinfo
  {pages} {184508} (\bibinfo {year} {2021})}\BibitemShut {NoStop}%
\bibitem [{\citenamefont {Kiessling}\ \emph {et~al.}(2025)\citenamefont
  {Kiessling}, \citenamefont {Altshuler},\ and\ \citenamefont
  {Yuzbashyan}}]{Kiessling2025Bounds}%
  \BibitemOpen
  \bibfield  {author} {\bibinfo {author} {\bibfnamefont {M.~K.-H.}\
  \bibnamefont {Kiessling}}, \bibinfo {author} {\bibfnamefont {B.~L.}\
  \bibnamefont {Altshuler}},\ and\ \bibinfo {author} {\bibfnamefont {E.~A.}\
  \bibnamefont {Yuzbashyan}},\ }\bibfield  {title} {\bibinfo {title} {Bounds on
  {$T_c$} in the {E}liashberg theory of superconductivity. {I}: The
  $\gamma$-model},\ }\href {https://doi.org/10.1007/s10955-025-03446-5}
  {\bibfield  {journal} {\bibinfo  {journal} {Journal of Statistical Physics}\
  }\textbf {\bibinfo {volume} {192}},\ \bibinfo {pages} {69} (\bibinfo {year}
  {2025})}\BibitemShut {NoStop}%
\bibitem [{\citenamefont {Patel}\ \emph {et~al.}(2018)\citenamefont {Patel},
  \citenamefont {McGreevy}, \citenamefont {Arovas},\ and\ \citenamefont
  {Sachdev}}]{Patel2018}%
  \BibitemOpen
  \bibfield  {author} {\bibinfo {author} {\bibfnamefont {A.~A.}\ \bibnamefont
  {Patel}}, \bibinfo {author} {\bibfnamefont {J.}~\bibnamefont {McGreevy}},
  \bibinfo {author} {\bibfnamefont {D.~P.}\ \bibnamefont {Arovas}},\ and\
  \bibinfo {author} {\bibfnamefont {S.}~\bibnamefont {Sachdev}},\ }\bibfield
  {title} {\bibinfo {title} {Magnetotransport in a model of a disordered
  strange metal},\ }\href {https://doi.org/10.1103/PhysRevX.8.021049}
  {\bibfield  {journal} {\bibinfo  {journal} {Phys. Rev. X}\ }\textbf {\bibinfo
  {volume} {8}},\ \bibinfo {pages} {021049} (\bibinfo {year}
  {2018})}\BibitemShut {NoStop}%
\bibitem [{Note1()}]{Note1}%
  \BibitemOpen
  \bibinfo {note} {Sampling the coupling constants from the real random
  Gaussian distribution (instead of the complex one) ensures the presence of
  anomalous averages below $T_c$ in the large-$N$ theory \cite
  {Esterlis2019Cooper}}\BibitemShut {NoStop}%
\bibitem [{Sup()}]{SupMat}%
  \BibitemOpen
  \href@noop {} {}\bibinfo {note} {See the Supplemental Material for further
  details on the derivations and numerical analysis in the main
  text.}\BibitemShut {Stop}%
\bibitem [{\citenamefont {Abrikosov}\ \emph {et~al.}(1965)\citenamefont
  {Abrikosov}, \citenamefont {Gor'kov},\ and\ \citenamefont
  {Dzyaloshinskii}}]{AGD}%
  \BibitemOpen
  \bibfield  {author} {\bibinfo {author} {\bibfnamefont {A.~A.}\ \bibnamefont
  {Abrikosov}}, \bibinfo {author} {\bibfnamefont {L.~P.}\ \bibnamefont
  {Gor'kov}},\ and\ \bibinfo {author} {\bibfnamefont {I.~Y.}\ \bibnamefont
  {Dzyaloshinskii}},\ }\href@noop {} {\emph {\bibinfo {title} {Quantum field
  theoretical methods in statistical physics}}},\ \bibinfo {edition} {2nd}\
  ed.\ (\bibinfo  {publisher} {Pergamon press Ltd., Oxford},\ \bibinfo {year}
  {1965})\BibitemShut {NoStop}%
\bibitem [{\citenamefont {Abrikosov}(1988)}]{Abrikosov1988Fundamentals}%
  \BibitemOpen
  \bibfield  {author} {\bibinfo {author} {\bibfnamefont {A.~A.}\ \bibnamefont
  {Abrikosov}},\ }\href@noop {} {\emph {\bibinfo {title} {Fundamentals of the
  Theory of Metals}}}\ (\bibinfo  {publisher} {North-Holland},\ \bibinfo {year}
  {1988})\BibitemShut {NoStop}%
\bibitem [{\citenamefont {Chubukov}\ and\ \citenamefont
  {Schmalian}(2005)}]{Chubukov2005Superconductivity}%
  \BibitemOpen
  \bibfield  {author} {\bibinfo {author} {\bibfnamefont {A.~V.}\ \bibnamefont
  {Chubukov}}\ and\ \bibinfo {author} {\bibfnamefont {J.}~\bibnamefont
  {Schmalian}},\ }\bibfield  {title} {\bibinfo {title} {Superconductivity due
  to massless boson exchange in the strong-coupling limit},\ }\href
  {https://doi.org/10.1103/PhysRevB.72.174520} {\bibfield  {journal} {\bibinfo
  {journal} {Phys. Rev. B}\ }\textbf {\bibinfo {volume} {72}},\ \bibinfo
  {pages} {174520} (\bibinfo {year} {2005})}\BibitemShut {NoStop}%
\bibitem [{Note2()}]{Note2}%
  \BibitemOpen
  \bibinfo {note} {The number $0.04$ agrees with the one predicted in the
  $\gamma =1$ model: $T_c \approx 0.25 {\protect \bar g}$~\cite {Wu2020gamma1},
  when accounting for the effective coupling constant $\protect \bar {g} =
  \varepsilon _F/(2\pi )$}\BibitemShut {NoStop}%
\bibitem [{\citenamefont {Marsiglio}\ and\ \citenamefont
  {Carbotte}(1991)}]{Marsiglio1991Gap}%
  \BibitemOpen
  \bibfield  {author} {\bibinfo {author} {\bibfnamefont {F.}~\bibnamefont
  {Marsiglio}}\ and\ \bibinfo {author} {\bibfnamefont {J.~P.}\ \bibnamefont
  {Carbotte}},\ }\bibfield  {title} {\bibinfo {title} {Gap function and density
  of states in the strong-coupling limit for an electron-boson system},\ }\href
  {https://doi.org/10.1103/PhysRevB.43.5355} {\bibfield  {journal} {\bibinfo
  {journal} {Phys. Rev. B}\ }\textbf {\bibinfo {volume} {43}},\ \bibinfo
  {pages} {5355} (\bibinfo {year} {1991})}\BibitemShut {NoStop}%
\bibitem [{Note3()}]{Note3}%
  \BibitemOpen
  \bibinfo {note} {For comparison, we demonstrate the unconstrained growth of
  $T_c$ in the strong-coupling limit of Migdal-Eliashberg theory in the absence
  of the self-energy of the boson in the Supplemental Material}\BibitemShut
  {NoStop}%
\bibitem [{\citenamefont {Heath}\ and\ \citenamefont
  {Boyack}(2025)}]{Heath2024}%
  \BibitemOpen
  \bibfield  {author} {\bibinfo {author} {\bibfnamefont {J.~T.}\ \bibnamefont
  {Heath}}\ and\ \bibinfo {author} {\bibfnamefont {R.}~\bibnamefont {Boyack}},\
  }\bibfield  {title} {\bibinfo {title} {Universal scaling relations in
  electron-phonon superconductors},\ }\href
  {https://doi.org/10.1103/PhysRevLett.134.216002} {\bibfield  {journal}
  {\bibinfo  {journal} {Phys. Rev. Lett.}\ }\textbf {\bibinfo {volume} {134}},\
  \bibinfo {pages} {216002} (\bibinfo {year} {2025})}\BibitemShut {NoStop}%
\bibitem [{\citenamefont {McMillan}(1968)}]{McMillan1968}%
  \BibitemOpen
  \bibfield  {author} {\bibinfo {author} {\bibfnamefont {W.~L.}\ \bibnamefont
  {McMillan}},\ }\bibfield  {title} {\bibinfo {title} {Transition temperature
  of strong-coupled superconductors},\ }\href
  {https://doi.org/10.1103/PhysRev.167.331} {\bibfield  {journal} {\bibinfo
  {journal} {Phys. Rev.}\ }\textbf {\bibinfo {volume} {167}},\ \bibinfo {pages}
  {331} (\bibinfo {year} {1968})}\BibitemShut {NoStop}%
\bibitem [{\citenamefont {Karakozov}\ \emph {et~al.}(1975)\citenamefont
  {Karakozov}, \citenamefont {Maksimov},\ and\ \citenamefont
  {Mashkov}}]{Karakozov1975}%
  \BibitemOpen
  \bibfield  {author} {\bibinfo {author} {\bibfnamefont {A.~E.}\ \bibnamefont
  {Karakozov}}, \bibinfo {author} {\bibfnamefont {E.~G.}\ \bibnamefont
  {Maksimov}},\ and\ \bibinfo {author} {\bibfnamefont {S.~A.}\ \bibnamefont
  {Mashkov}},\ }\bibfield  {title} {\bibinfo {title} {Effect of the frequency
  dependence of the electron-phonon interaction spectral function on the
  thermodynamic properties of superconductors},\ }\href
  {http://www.jetp.ras.ru/cgi-bin/e/index/e/41/5/p971?a=list} {\bibfield
  {journal} {\bibinfo  {journal} {Sov. Phys. JETP}\ }\textbf {\bibinfo {volume}
  {41}},\ \bibinfo {pages} {971} (\bibinfo {year} {1975})}\BibitemShut
  {NoStop}%
\bibitem [{\citenamefont {Son}(1999)}]{Son1999Superconductivity}%
  \BibitemOpen
  \bibfield  {author} {\bibinfo {author} {\bibfnamefont {D.~T.}\ \bibnamefont
  {Son}},\ }\bibfield  {title} {\bibinfo {title} {Superconductivity by
  long-range color magnetic interaction in high-density quark matter},\ }\href
  {https://doi.org/10.1103/PhysRevD.59.094019} {\bibfield  {journal} {\bibinfo
  {journal} {Phys. Rev. D}\ }\textbf {\bibinfo {volume} {59}},\ \bibinfo
  {pages} {094019} (\bibinfo {year} {1999})}\BibitemShut {NoStop}%
\bibitem [{\citenamefont {Sch\"afer}\ and\ \citenamefont
  {Wilczek}(1999)}]{Wilczek1999}%
  \BibitemOpen
  \bibfield  {author} {\bibinfo {author} {\bibfnamefont {T.}~\bibnamefont
  {Sch\"afer}}\ and\ \bibinfo {author} {\bibfnamefont {F.}~\bibnamefont
  {Wilczek}},\ }\bibfield  {title} {\bibinfo {title} {Superconductivity from
  perturbative one-gluon exchange in high density quark matter},\ }\href
  {https://doi.org/10.1103/PhysRevD.60.114033} {\bibfield  {journal} {\bibinfo
  {journal} {Phys. Rev. D}\ }\textbf {\bibinfo {volume} {60}},\ \bibinfo
  {pages} {114033} (\bibinfo {year} {1999})}\BibitemShut {NoStop}%
\bibitem [{\citenamefont {Pisarski}\ and\ \citenamefont
  {Rischke}(2000)}]{Pisarski2000}%
  \BibitemOpen
  \bibfield  {author} {\bibinfo {author} {\bibfnamefont {R.~D.}\ \bibnamefont
  {Pisarski}}\ and\ \bibinfo {author} {\bibfnamefont {D.~H.}\ \bibnamefont
  {Rischke}},\ }\bibfield  {title} {\bibinfo {title} {Gaps and critical
  temperature for color superconductivity},\ }\href
  {https://doi.org/10.1103/PhysRevD.61.051501} {\bibfield  {journal} {\bibinfo
  {journal} {Phys. Rev. D}\ }\textbf {\bibinfo {volume} {61}},\ \bibinfo
  {pages} {051501} (\bibinfo {year} {2000})}\BibitemShut {NoStop}%
\bibitem [{\citenamefont {Metlitski}\ \emph {et~al.}(2015)\citenamefont
  {Metlitski}, \citenamefont {Mross}, \citenamefont {Sachdev},\ and\
  \citenamefont {Senthil}}]{Metlitski2015Cooper}%
  \BibitemOpen
  \bibfield  {author} {\bibinfo {author} {\bibfnamefont {M.~A.}\ \bibnamefont
  {Metlitski}}, \bibinfo {author} {\bibfnamefont {D.~F.}\ \bibnamefont
  {Mross}}, \bibinfo {author} {\bibfnamefont {S.}~\bibnamefont {Sachdev}},\
  and\ \bibinfo {author} {\bibfnamefont {T.}~\bibnamefont {Senthil}},\
  }\bibfield  {title} {\bibinfo {title} {Cooper pairing in non-{F}ermi
  liquids},\ }\href {https://doi.org/10.1103/PhysRevB.91.115111} {\bibfield
  {journal} {\bibinfo  {journal} {Phys. Rev. B}\ }\textbf {\bibinfo {volume}
  {91}},\ \bibinfo {pages} {115111} (\bibinfo {year} {2015})}\BibitemShut
  {NoStop}%
\bibitem [{\citenamefont {Gnezdilov}\ and\ \citenamefont
  {Wang}(2022)}]{Gnezdilov2022Solvable}%
  \BibitemOpen
  \bibfield  {author} {\bibinfo {author} {\bibfnamefont {N.~V.}\ \bibnamefont
  {Gnezdilov}}\ and\ \bibinfo {author} {\bibfnamefont {Y.}~\bibnamefont
  {Wang}},\ }\bibfield  {title} {\bibinfo {title} {Solvable model for a
  charge-$4e$ superconductor},\ }\href
  {https://doi.org/10.1103/PhysRevB.106.094508} {\bibfield  {journal} {\bibinfo
   {journal} {Phys. Rev. B}\ }\textbf {\bibinfo {volume} {106}},\ \bibinfo
  {pages} {094508} (\bibinfo {year} {2022})}\BibitemShut {NoStop}%
\bibitem [{\citenamefont {Michon}\ \emph {et~al.}(2023)\citenamefont {Michon},
  \citenamefont {Berthod}, \citenamefont {Rischau}, \citenamefont {Ataei},
  \citenamefont {Chen}, \citenamefont {Komiya}, \citenamefont {Ono},
  \citenamefont {Taillefer}, \citenamefont {van~der Marel},\ and\ \citenamefont
  {Georges}}]{Michon2023Reconciling}%
  \BibitemOpen
  \bibfield  {author} {\bibinfo {author} {\bibfnamefont {B.}~\bibnamefont
  {Michon}}, \bibinfo {author} {\bibfnamefont {C.}~\bibnamefont {Berthod}},
  \bibinfo {author} {\bibfnamefont {C.~W.}\ \bibnamefont {Rischau}}, \bibinfo
  {author} {\bibfnamefont {A.}~\bibnamefont {Ataei}}, \bibinfo {author}
  {\bibfnamefont {L.}~\bibnamefont {Chen}}, \bibinfo {author} {\bibfnamefont
  {S.}~\bibnamefont {Komiya}}, \bibinfo {author} {\bibfnamefont
  {S.}~\bibnamefont {Ono}}, \bibinfo {author} {\bibfnamefont {L.}~\bibnamefont
  {Taillefer}}, \bibinfo {author} {\bibfnamefont {D.}~\bibnamefont {van~der
  Marel}},\ and\ \bibinfo {author} {\bibfnamefont {A.}~\bibnamefont
  {Georges}},\ }\bibfield  {title} {\bibinfo {title} {Reconciling scaling of
  the optical conductivity of cuprate superconductors with {P}lanckian
  resistivity and specific heat},\ }\href
  {https://doi.org/10.1038/s41467-023-38762-5} {\bibfield  {journal} {\bibinfo
  {journal} {Nature Communications}\ }\textbf {\bibinfo {volume} {14}},\
  \bibinfo {pages} {3033} (\bibinfo {year} {2023})}\BibitemShut {NoStop}%
\bibitem [{\citenamefont {Badoux}\ \emph {et~al.}(2016)\citenamefont {Badoux},
  \citenamefont {Afshar}, \citenamefont {Michon}, \citenamefont {Ouellet},
  \citenamefont {Fortier}, \citenamefont {LeBoeuf}, \citenamefont {Croft},
  \citenamefont {Lester}, \citenamefont {Hayden}, \citenamefont {Takagi},
  \citenamefont {Yamada}, \citenamefont {Graf}, \citenamefont
  {Doiron-Leyraud},\ and\ \citenamefont {Taillefer}}]{Badoux2016}%
  \BibitemOpen
  \bibfield  {author} {\bibinfo {author} {\bibfnamefont {S.}~\bibnamefont
  {Badoux}}, \bibinfo {author} {\bibfnamefont {S.~A.~A.}\ \bibnamefont
  {Afshar}}, \bibinfo {author} {\bibfnamefont {B.}~\bibnamefont {Michon}},
  \bibinfo {author} {\bibfnamefont {A.}~\bibnamefont {Ouellet}}, \bibinfo
  {author} {\bibfnamefont {S.}~\bibnamefont {Fortier}}, \bibinfo {author}
  {\bibfnamefont {D.}~\bibnamefont {LeBoeuf}}, \bibinfo {author} {\bibfnamefont
  {T.~P.}\ \bibnamefont {Croft}}, \bibinfo {author} {\bibfnamefont
  {C.}~\bibnamefont {Lester}}, \bibinfo {author} {\bibfnamefont {S.~M.}\
  \bibnamefont {Hayden}}, \bibinfo {author} {\bibfnamefont {H.}~\bibnamefont
  {Takagi}}, \bibinfo {author} {\bibfnamefont {K.}~\bibnamefont {Yamada}},
  \bibinfo {author} {\bibfnamefont {D.}~\bibnamefont {Graf}}, \bibinfo {author}
  {\bibfnamefont {N.}~\bibnamefont {Doiron-Leyraud}},\ and\ \bibinfo {author}
  {\bibfnamefont {L.}~\bibnamefont {Taillefer}},\ }\bibfield  {title} {\bibinfo
  {title} {Critical doping for the onset of {F}ermi-surface reconstruction by
  charge-density-wave order in the cuprate superconductor
  {${\mathrm{La}}_{2\ensuremath{-}x}{\mathrm{Sr}}_{x}{\mathrm{CuO}}_{4}$}},\
  }\href {https://doi.org/10.1103/PhysRevX.6.021004} {\bibfield  {journal}
  {\bibinfo  {journal} {Phys. Rev. X}\ }\textbf {\bibinfo {volume} {6}},\
  \bibinfo {pages} {021004} (\bibinfo {year} {2016})}\BibitemShut {NoStop}%
\end{thebibliography}%

\end{document}


\title{Supplemental Material for ``Upper bound on $T_c$ in a strongly coupled electron-boson superconductor''}

\author{Nikolay V. Gnezdilov}
\email{Nikolay.Gnezdilov@dartmouth.edu}

\author{Rufus Boyack}
\email{Rufus.Boyack@dartmouth.edu}
\affiliation{Department of Physics and Astronomy, Dartmouth College, Hanover, New Hampshire 03755, USA}

\maketitle

In this Supplemental Material, we (i) provide the derivation of the saddle-point equations for the $2$d Y-SYK model; (ii) obtain the linearized gap equation from the Y-SYK saddle-point equations; (iii) explain the numerical method we use to solve for $T_c$; and (iv) derive $T_c$ in the strong-coupling limit in the absence of the bosonic self-energy within the same framework as used in the main text.

\section{I. Derivation of the saddle-point equations for the 2d Y-SYK model}

The action describing $N$ electrons interacting with $N$ bosons in $(2+1)$d through a spatially disordered Yukawa-SYK interaction reads $S = S_{\psi} + S_{\phi} + S_{\psi\phi}$, where the individual contributions are 
\begin{align}
    S_{\psi} & = \int_x \, \sum_{i=1}^N \sum_{\sigma=\uparrow\downarrow} \psi_{i\sigma}^\dag(x) \left(\partial_\tau - \tfrac{\nabla_{\bf r}^2}{2m_0}-\mu\right) \psi_{i\sigma}(x); \quad 
    S_{\phi}  =  \int_x \, \sum_{i=1}^N \frac{1}{2} \phi_i(x) (-\partial_\tau^2 -c^2 \nabla_{\bf r}^2 + \epsilon^2) \phi_i(x), \label{S_0} \\
    S_{\psi\phi} & =  \frac{1}{N} \sum_{ijl=1}^N   \int_x  g_{ijl}({\bf r}) \sum_{\sigma=\uparrow\downarrow}\psi_{i\sigma}^\dag(x) \psi_{j\sigma}(x) \phi_l(x) = \frac{1}{N} \sum_{ijl=1}^N   \int_x g_{ijl}({\bf r}) {\cal O}_{ij}(x)\phi_l(x), \label{S_YSYK} 
\end{align}
where $\psi_{i\sigma}$, $\psi_{i\sigma}^\dag$ are fermions with spin $\sigma=\uparrow\downarrow$, mass $m_0$, and a chemical potential $\mu$; $\phi_i$ are bosons with mass $\epsilon$, and $c$ is the speed of sound. The variable $x$ denotes $x = (\tau,{\bf r})$ and the integration notation is $\int_x\equiv\int_0^{\beta}d\tau\int d^2r$, where $\beta = 1/T$ is the inverse temperature. For convenience, we define ${\cal O}_{ij}(x) = \sum_{\sigma} \psi_{i\sigma}^\dag(x) \psi_{j\sigma}(x)$.
To allow for anomalous propagators in the large-$N$ limit~\cite{Esterlis2019Cooper}, we draw the coupling constants $g_{ijl}$ in the Yukawa-SYK term \eqref{S_YSYK} independently from a Gaussian orthogonal ensemble with zero mean and finite variance $\langle g_{ijl}({\bf r}) g_{ijl} ({\bf r}')\rangle = 2 g^2 \delta({\bf r}-{\bf r}')$.

Performing the disorder average gives
\begin{equation}
\langle e^{-S_{\psi\phi}}\rangle \!=\! \exp\!\bigg[\! \frac{g^2}{4N^2} \!\int_{x,x'}\!\! \delta({\bf r}-{\bf r}') \!\sum_{l}^N\!\phi_l(x) \phi_l(x')   \!\sum_{ij}^N({\cal O}^\dag_{ij}(x) {\cal O}^\dag_{ij}(x') \!+\! {\cal O}^\dag_{ij}(x) {\cal O}_{ij}(x') \!+\! {\cal O}_{ij}(x) {\cal O}^\dag_{ij}(x') \!+\! {\cal O}_{ij}(x) {\cal O}_{ij}(x') )\!\bigg]. \label{dav1}
\end{equation}
To decouple the interactions in Eq.~\eqref{dav1}, we use the following resolutions of unity in terms of auxiliary bilocal fields:
\begin{align} 
    {\cal I} &= \int \! \mathcal{D}\Sigma\mathcal{D}G \exp\bigg[  \int_{x,x'}  \sum_{\sigma=\uparrow\downarrow} \Sigma_{\sigma}(x-x') (N G_{\sigma}(x'\!-x) - \sum_{i=1}^N \psi^\dag_{i\sigma}(x)\psi_{i\sigma}(x') ) \bigg], \label{G_def}\\ 
    {\cal I} &=  \int \! \mathcal{D}\Phi\mathcal{D}F^+ \exp\bigg[  \int_{x,x'}\Phi_{\uparrow\downarrow}(x-x') (N F^+_{\downarrow\uparrow}(x'\!-x) - \sum_{i=1}^N \psi^\dag_{i\uparrow}(x)\psi^\dag_{i\downarrow}(x') ) \bigg], \label{F+_def}\\ 
    {\cal I}  &= \int \! \mathcal{D}\Phi^+ \mathcal{D}F \exp\bigg[  \int_{x,x'} \Phi^+_{\downarrow\uparrow}(x-x') (N F_{\uparrow\downarrow}(x'\!-x) - \sum_{i=1}^N \psi_{i\downarrow}(x)\psi_{i\uparrow}(x') ) \bigg],  \label{F_def} \\
    {\cal I}  &= \int \! \mathcal{D}\Pi\mathcal{D}D \exp\bigg[ - \frac{1}{2}\int_{x,x'}  \Pi(x-x') (N D(x'\!-x) - \sum_{i=1}^N \phi_{i}(x)\phi_{i}(x') ) \bigg]. \label{D_def}
\end{align}  
We assume that the Cooper pairing occurs in the spin-singlet channel and the normal components of the fermionic Green's function are spin diagonal. Using Eqs.~(\ref{G_def}-\ref{F_def}), we have 
\begin{align} 
   \frac{1}{N} \sum_{ij}^N{\cal O}_{ij}(x) {\cal O}^\dag_{ij}(x') &= \frac{1}{N} \sum_{ij}^N \sum_{\sigma\sigma'}\psi_{i\sigma}^\dag(x) \psi_{j\sigma}(x) \psi_{j\sigma'}^\dag(x') \psi_{i\sigma'}(x')  
   = - N \sum_{\sigma=\uparrow\downarrow}  G_{\sigma}(x-x') G_{\sigma}(x'\!-x), \label{OOdag}\\ 
   \frac{1}{N} \sum_{ij}^N{\cal O}_{ij}^\dag(x) {\cal O}_{ij}(x') &= \frac{1}{N} \sum_{ij}^N \sum_{\sigma\sigma'}\psi_{j\sigma}^\dag(x) \psi_{i\sigma}(x) \psi_{i\sigma'}^\dag(x') \psi_{j\sigma'}(x')  
   = - N \sum_{\sigma=\uparrow\downarrow}  G_{\sigma}(x-x') G_{\sigma}(x'\!-x), \label{OdagO}\\ \nonumber
   \frac{1}{N} \sum_{ij}^N{\cal O}_{ij}(x) {\cal O}_{ij}(x') &= \frac{1}{N} \sum_{ij}^N (\psi_{i\uparrow}^\dag(x) \psi_{j\uparrow}(x) \psi_{i\downarrow}^\dag(x') \psi_{j\downarrow}(x') + \psi_{i\downarrow}^\dag(x) \psi_{j\downarrow}(x) \psi_{i\uparrow}^\dag(x') \psi_{j\uparrow}(x'))\\ 
   &= N(F_{\uparrow\downarrow}(x-x')F^+_{\downarrow\uparrow}(x'\!-x)+ F^+_{\downarrow\uparrow} (x-x')F_{\uparrow\downarrow}(x'\!-x)), \label{OO}\\ \nonumber
   \frac{1}{N} \sum_{ij}^N{\cal O}_{ij}^\dag(x) {\cal O}_{ij}^\dag(x') &= \frac{1}{N} \sum_{ij}^N ( \psi_{j\uparrow}^\dag(x) \psi_{i\uparrow}(x) \psi_{j\downarrow}^\dag(x') \psi_{i\downarrow}(x') + \psi_{j\downarrow}^\dag(x) \psi_{i\downarrow}(x) \psi_{j\uparrow}^\dag(x') \psi_{i\uparrow}(x') ) \\ 
   &= N(F_{\uparrow\downarrow}(x-x')F^+_{\downarrow\uparrow}(x'\!-x)+ F^+_{\downarrow\uparrow} (x-x')F_{\uparrow\downarrow}(x'\!-x)). \label{OdagOdag}
\end{align}
For a spin-diagonal normal channel and spin-singlet pairing, we set $\Sigma_{\uparrow\uparrow} = \Sigma_{\downarrow\downarrow}=\Sigma$, $G_{\uparrow\uparrow} = G_{\downarrow\downarrow}=G$, $\Phi_{\uparrow\downarrow}=-\Phi_{\downarrow\uparrow}=\Phi$, $\Phi^+_{\downarrow\uparrow}=-\Phi^+_{\uparrow\downarrow}=\Phi^*$, $F_{\uparrow\downarrow}=-F_{\downarrow\uparrow}=F$, $F^+_{\downarrow\uparrow}=-F^+_{\uparrow\downarrow}=F^*$.
From Eqs.~(\ref{S_0}-\ref{dav1},\ref{OOdag}-\ref{OdagOdag}), we note that the action $S$ is quadratic in the fermionic and bosonic fields, and thus they can be integrated out exactly, leading to 
\begin{align}
S &= S_{\psi\Sigma} + S_{\phi\Pi} + S_{\Sigma G} + S_{\Pi D} + S_{DG}, \label{S_eff}\\ 
S_{\psi\Sigma} &= -N \, {\rm Tr}\log  
  \begin{pmatrix}
  (\partial_\tau - \frac{\nabla_{\bf r}^2}{2m_{0}} - \mu)\delta(x-x')  + \Sigma(x-x') & \Phi(x-x') \\ \Phi^*(x-x') &  (\partial_\tau + \frac{\nabla_{\bf r}^2}{2m_{0}} + \mu)\delta(x-x')  - \Sigma(x'\!-x) 
  \end{pmatrix},\\ 
   S_{\phi\Pi} &= \frac{N}{2} \, {\rm Tr}\log\big((-\partial_\tau^2 - c^2 \nabla_{\bf r}^2 +\epsilon^2)\delta(x-x') - \Pi(x-x')\big) ,\\ 
  S_{\Sigma G} &= -N \int_{x,x'} ( 2  \Sigma(x-x') G(x'\!-x) + \Phi(x-x') F^*(x'\!-x) + \Phi^*(x-x') F(x'\!-x) ),\\
  S_{\Pi D} &= \frac{N}{2} \int_{x,x'} \Pi(x-x')  D(x'\!-x), \\ 
  S_{DG} &= \frac{N g^2}{2} \int_{x,x'}  \delta(\mathbf{r}-\mathbf{r}^\prime) ( 2 G(x-x') G(x'\!-x)  - F(x-x')F^*(x'\!-x) - F^*(x-x')F(x'\!-x) ) D(x'\!-x). \label{S_DG}
\end{align}

In the large-$N$ limit, we obtain the saddle-point equations for $G, \Sigma, F, \Phi, \Pi, D$. To do so, we set the variation of the effective action \eqref{S_eff} with respect to the bilocal fields equal to zero. Performing these variations of $S$ gives
\begin{align}
    G(i\omega_n, {\bf k}) &= \frac{i\omega_n + \xi_{-\bf k}+ \Sigma(-i\omega_n)}{(i\omega_n - \xi_{\bf k} - \Sigma(i\omega_n))(i\omega_n + \xi_{-\bf k}+ \Sigma(-i\omega_n))-|\Phi(i\omega_n)|^2}, \\
    F(i\omega_n, {\bf k}) &= \frac{\Phi(i\omega_n)}{(i\omega_n - \xi_{\bf k} - \Sigma(i\omega_n))(i\omega_n + \xi_{-\bf k}+ \Sigma(-i\omega_n))-|\Phi(i\omega_n)|^2}, \\
     D(i\Omega_m, {\bf q}) &= \frac{1}{\Omega_m^2 + c^2q^2 + \epsilon^2 - \Pi(i \Omega_m)}, \\
    \Sigma(i\omega_n) &= g^2 T \sum_{\omega_n} \int \! D(i\omega_n-i\omega_{n'},{\bf q}) G(i\omega_{n'},{\bf k}') \frac{d^2q}{(2\pi)^2}\frac{d^2k'}{(2\pi)^2}, \label{Sigma_eq}  \\
    \Phi(i\omega_n) &= -g^2 T \sum_{\omega_{n'}} \int \! D(i\omega_n-i\omega_{n'},{\bf q}) F(i\omega_{n'},{\bf k}') \frac{d^2q}{(2\pi)^2}\frac{d^2k'}{(2\pi)^2},\label{Phi_eq} \\ \nonumber
   \Pi(i\Omega_m) &= -g^2 T \sum_{\omega_n} \int \!  \bigg( 2 G(i\omega_n, {\bf k}) G(i\omega_n+i\Omega_m, {\bf k}') \\ &\quad - F(i\omega_n, {\bf k})F^*(i\omega_n+i\Omega_m, {\bf k}') - F^* (i\omega_n, {\bf k})F(i\omega_n+i\Omega_m, {\bf k}') \bigg) \frac{d^2k}{(2\pi)^2}\frac{d^2k'}{(2\pi)^2}, \label{Pi_eq}
\end{align}
where $\omega_n = (2n+1)\pi T$ and $\Omega_m = 2m\pi T$ are fermionic and bosonic Matsubara frequencies, respectively. The momentum integrals in Eqs.~(\ref{Sigma_eq}-\ref{Pi_eq}) factorize because of the spatial delta function in Eq.~\eqref{S_DG}.

\section{II. Derivation of the linearized gap equation from the Y-SYK saddle-point equations}

In the large-$N$ limit, the linearized equation for the superconducting order parameter $\Phi$ at $T=T_c$ is
\begin{equation}
    \Phi(i\omega_n) = g^2 T_c \sum_{\omega_{n'}} \int \! D(i\omega_n-i\omega_{n'},{\bf q}) G(i\omega_{n'},{\bf k}') G(-i\omega_{n'},-{\bf k}') \Phi(i\omega_{n'}) \frac{d^2q}{(2\pi)^2}\frac{d^2k'}{(2\pi)^2}. \label{Phi_lin_eq}
\end{equation}

Following the main text and noting the factorization of the momenta integrals in Eq.~\eqref{Phi_lin_eq}, we have
\begin{equation}
    \int \! D(i\Omega_m, {\bf q}) \frac{d^2q}{(2\pi)^2} = \int_0^{\omega_D/c} \frac{1}{\Omega_m^2 + c^2q^2 - \Pi(i \Omega_m)} \frac{qdq}{2\pi} = \frac{1}{4\pi c^2}\log \!\left[1\!+\!\frac{1}{(\frac{\Omega_m}{\omega_D})^2 \! + \! 2 \pi \lambda\frac{|\Omega_m|}{\varepsilon_F}} \right] = \frac{L(i\Omega_m)}{4\pi c^2}, \label{D_eff}
\end{equation}
where $\lambda = \nu g^2/(4\pi c^2)$ is the dimensionless electron-boson coupling constant, $\nu$ is the $2$d electronic density of states, $c$ is the speed of sound, the bosonic mass $\epsilon$ is set to zero, and $\Pi$ is the bosonic self-energy in the normal state. We introduced an analog of the Debye frequency $\omega_D$ as a UV cutoff for the bosonic propagator (\ref{D_eff}), which is equivalent to performing Pauli-Villars regularization~\cite{Esterlis2021Large}. Using Eq.~\eqref{Pi_eq},  $\Pi(i\Omega_m) = - 2 g^2 T \sum_{\omega_n} {\cal G}(i\omega_n){\cal G}(i\omega_n+i\Omega_m)$, where ${\cal G}(i\omega_n)=\int \! G(i\omega_n, {\bf k}) \frac{d^2k}{(2\pi)^2}=-i\pi \nu {\rm sgn}(\omega_n)$ is the electronic propagator integrated over momentum. Substituting the effective propagator ${\cal G}$ in the bosonic self-energy, we have $\Pi(i\Omega_m)=2\pi^2 \nu^2 g^2 T(-2|m| + \sum_{\omega_n}1) = - 2\pi \nu^2 g^2 |\Omega_m|$, where we used Riemann’s zeta function $\zeta(z)$ to regularize the sum: $\sum_{\omega_n}1= 1+2\zeta(0)=0$. Expressing $g^2$ via dimensionless electron-boson coupling $\lambda$ and assuming that the Debye momentum is of the order of the Fermi momentum~\cite{AGD,Abrikosov1988Fundamentals}, we obtain the contribution of the bosonic self-energy in Eq.~(\ref{D_eff}). Alternatively, one can obtain the bosonic self-energy in the time domain. Performing the Fourier transform of the effective electronic propagator, we get $\mathcal{G}(\tau)= -\pi\nu T/{\sin(\pi\tau T)}$, where $-\beta<\tau<\beta$, which agrees with Ref.~\cite{Patel2018} (up to the redefinition of the density of states). Writing Eq.~(\ref{Pi_eq}) in the normal state in the time domain, we have $\Pi(\tau)=-2g^{2}\mathcal{G}(\tau)\mathcal{G}(-\tau)=2(\pi \nu  T)^2/\sin^2(\pi\tau T)$. The latter expression agrees with the Fourier transform of the bosonic self-energy in the frequency domain that we derived above. Importantly, since the propagators are exact in the large-$N$ limit and, hence, non-perturbative in the coupling strength, the expression for the bosonic self-energy is valid at any coupling and does not have a static contribution, in contrast to the electron-phonon physics~\cite{Chubukov2020Eliashberg, AGD}.

Next, we compute 
\begin{equation}
    \int \! G(i\omega_n,{\bf k}) G(-i\omega_{n},-{\bf k}) \frac{d^2k}{(2\pi)^2} = \nu \int_{-\infty}^{\infty} \frac{d\xi}{\xi^2 +\omega^2_n Z(i\omega_n)^2} = \frac{\nu \pi}{|\omega_n|Z(i\omega_n)} \label{GG},
\end{equation}
where we used that $\Sigma(i\omega_n) = i\omega_n(1-Z(i\omega_n))$ with $Z(i\omega_n)$ being an even function of frequency.

Substituting \eqref{D_eff} and \eqref{GG} into Eq.~\eqref{Phi_lin_eq} and defining the gap function $\Delta(i\omega_n)=\Phi(i\omega_n)/Z(i\omega_n)$, we obtain the linearized gap equation:
\begin{equation}
Z(i\omega_n) \Delta(i\omega_n) = \pi \lambda T_c \sum_{\omega_{n'}} \frac{\Delta(i\omega_{n'})}{|\omega_{n'}|} L(i\omega_n-i\omega_{n'}). 
\label{ZDelta} 
\end{equation}
The renormalization function follows from the normal-state self-energy: 
\begin{align}  \nonumber
Z(i\omega_n) =& 1+\frac{i\Sigma(i\omega_n)}{\omega_n} =1 +\frac{i g^2 T_c }{\omega_n}\sum_{\omega_n} \int \! D(i\omega_n-i\omega_{n'},{\bf q}) G(i\omega_{n'},{\bf k}') \frac{d^2q}{(2\pi)^2}\frac{d^2k'}{(2\pi)^2} \\=&1 + \pi \lambda  \frac{T_c}{\omega_n} \sum_{\omega_{n'}} {\rm sgn}(\omega_{n'}) L(i\omega_n-i\omega_{n'}). \label{Z}
\end{align}
We then substitute the renormalization function from Eq.~\eqref{Z} into Eq.~\eqref{ZDelta} and derive the linearized gap equation from the main text:
\begin{equation}
\Delta(i\omega_n) = \pi \lambda  T_c \sum_{\omega_{n'}} \bigg[\Delta(i\omega_{n'}) - \frac{\omega_{n'}}{\omega_n} \Delta(i\omega_n) \bigg]
\frac{L(i\omega_n-i\omega_{n'})}{|\omega_{n'}|}. \label{gap_eq} 
\end{equation}

Since the resonant term ($n=n'$) in Eq.~(\ref{gap_eq}) does not contribute to $T_c$, we exclude it and write the linearized gap equation for positive frequencies:
\begin{equation} 
z(i\omega_n) \Delta(i\omega_n) = \pi \lambda T_c \sum_{n'=0}^{+\infty} \frac{\Delta(i\omega_{n'})}{\omega_{n'}}  \bigg[(1\!-\!\delta_{nn'}) L(i\omega_n\!-\!i\omega_{n'}) \!+\! L(i\omega_n\!+\!i\omega_{n'})  \bigg], \label{gap_eq_pos}
\end{equation}
where the new renormalization function $z(i\omega_n)$ is analogous to Eq.~(\ref{Z}) but with $n'\!\neq n$ in the summation. After the exclusion of the resonant term, we change the summation index from $n'$ to $m = n-n'$, split the contributions to the series for positive and negative frequency $\omega_n$, and use the evenness of the bosonic propagator to obtain 
\begin{equation}
z(i\omega_n) = 1 + 2\pi\lambda  \,\frac{T_c}{\omega_n} \sum_{m=1}^n L(i\Omega_m). 
\end{equation}

\section{III. Numerical solution of the pairing problem}

\begin{figure}[t!]
\center
\includegraphics[width=0.37\linewidth]{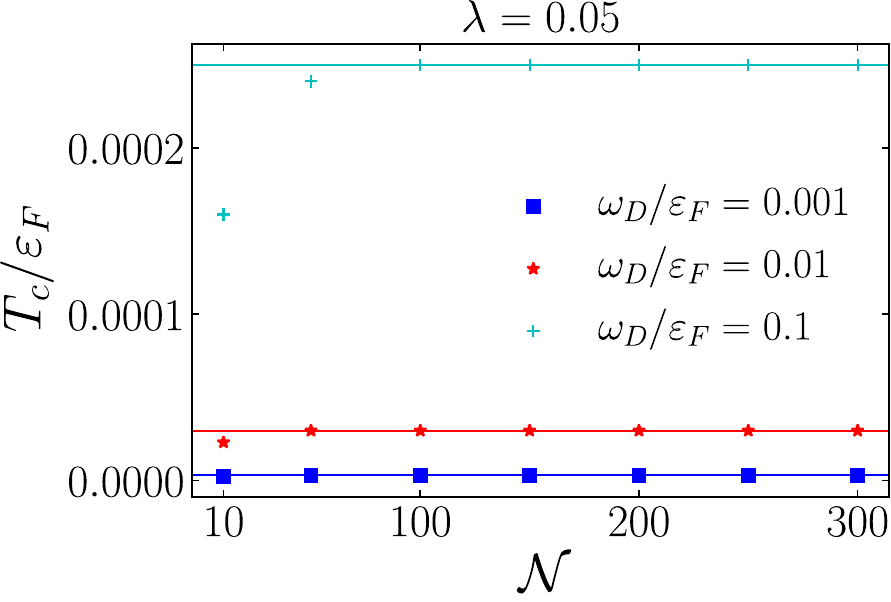}  
\hspace{0.1\linewidth}
\includegraphics[width=0.37\linewidth]{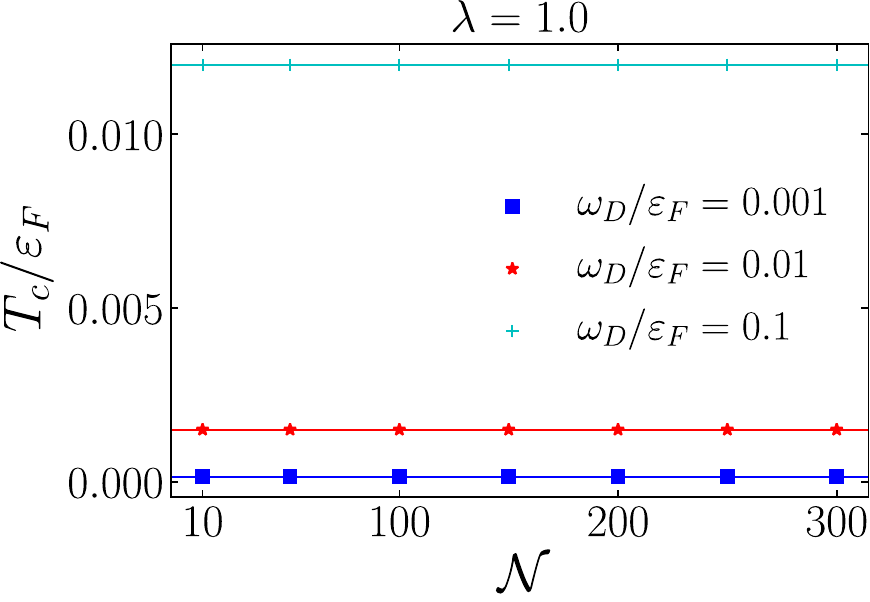} 
\caption{\small \label{fig:S1} {\bf Convergence of $T_c$ in the adiabatic regime} ($\omega_D/\varepsilon_F\ll 1$): $T_c$ as a function of the number of Matsubara frequencies ${\cal N}$ used to solve the pairing problem for $\lambda=0.05$ ({\it left}) and $\lambda=1$ ({\it right}). The color scheme is the same as in the main text.}
\end{figure}

\begin{figure}[t!]
\center
\includegraphics[width=0.37\linewidth]{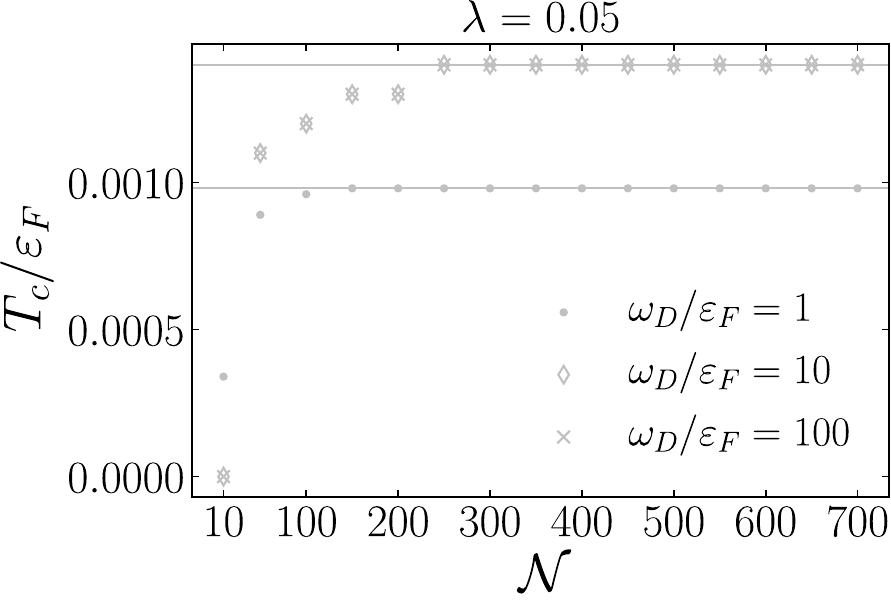}  
\hspace{0.1\linewidth}
\includegraphics[width=0.37\linewidth]{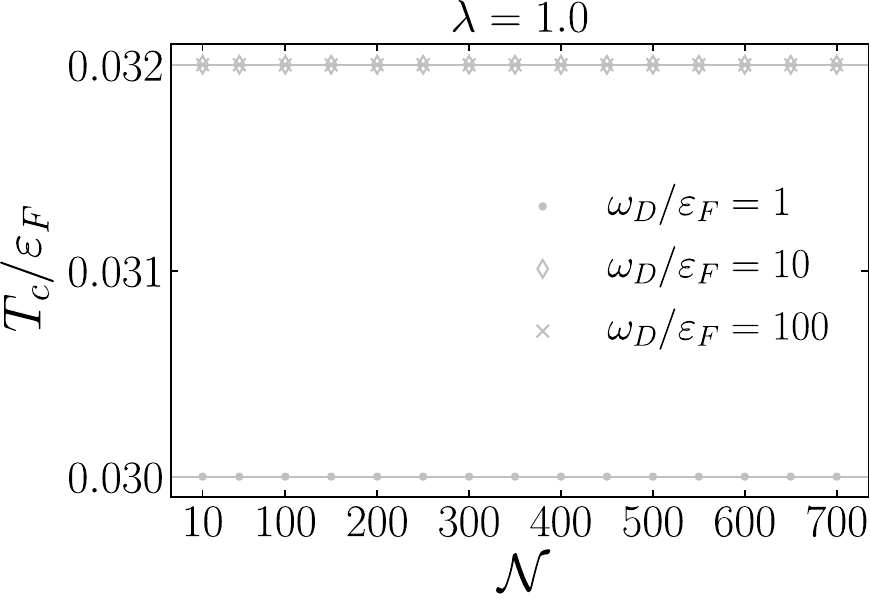} 
\caption{\small \label{fig:S2} {\bf Convergence of $T_c$ in the non-adiabatic regime} ($\omega_D/\varepsilon_F\geq 1$): $T_c$ as a function of the number of Matsubara frequencies ${\cal N}$ used to solve the pairing problem for $\lambda=0.05$ ({\it left}) and $\lambda=1$ ({\it right}). The color scheme is the same as in the main text.}
\end{figure}

We find the critical temperature by solving the characteristic equation (13) with the matrix (14) in the main text. For a given adiabaticity ratio $\omega_D/\varepsilon_F$ and coupling strength $\lambda$, we guess the bounds of the temperature interval $[T_{\rm min}, T_{\rm max}]$, over which we look for a zero of $F(T) = {\rm det}[{\cal M}(T)-{\cal I}]$, where $T\in[T_{\rm min}, T_{\rm max}]$, assuming that $F(T)$ is a continuous monotonic function on $[T_{\rm min}, T_{\rm max}]$. The matrix ${\cal M}$ is ${\cal N}\times{\cal N}$, where ${\cal N}$ is the number of positive Matsubara frequencies. To ensure that a zero of $F(T)$ is within $[T_{\rm min}, T_{\rm max}]$, we check that $F(T_{\rm min}) F(T_{\rm max})<0$. We divide the interval into $M=91$ equidistant points, including the boundaries, so that $T_0 = T_{\rm min}$ and $T_{M-1}=T_{\rm max}$. Starting from $t=0$, we compute $f_t = F(T_t) F(T_{t+1})$ until we find $f_t < 0$ in some step $t=t_c$. Then, $T_c = T_{t_c}$ if $|F(T_{t_c})|<|F(T_{t_c+1})|$ and $T_c = T_{t_c+1}$ if $|F(T_{t_c})|>|F(T_{t_c+1})|$. For example, for $\omega_D/\varepsilon_F = 0.001$ and $\lambda = 1$, we use ${\cal N}=100$ and choose $T_{\rm min}=10^{-4}\,\varepsilon_F$ and $T_{\rm max} = 10^{-3}\,\varepsilon_F$ with a step of $10^{-5}\varepsilon_F$, which gives us $T_c = 1.5 \cdot 10^{-4}\, \varepsilon_F = 0.15 \, \omega_D$.

The number of Matsubara frequencies ${\cal N}$ sufficient for convergence of the method depends on the coupling strength and the adiabaticity ratio.  At weak coupling, we generally deal with low temperatures~\cite{McMillan1968} and, hence, we need to account for more Matsubara frequencies than in the intermediate/strong coupling case for the $T_c$ solution to converge. 

We show the convergence of $T_c$ for $\omega_D/\varepsilon_F \ll 1$ by plotting $T_c$ in units of $\varepsilon_F$ in Fig.~{\ref{fig:S1}} for $\lambda=0.05$ (left panel) and $\lambda=1$ (right panel). For $\lambda=0.05$ (the lowest value of $\lambda$ we consider), $T_c$ converges for ${\cal N}=100$ frequencies for $\omega_D/\varepsilon_F = 0.1$ and with ${\cal N}=50$ for $\omega_D/\varepsilon_F = 0.01, 0.001$. For $\lambda=1$, ${\cal N}=10$ is sufficient for $T_c$ to converge. For our $T_c$ calculation shown in the main text in the adiabatic regime ($\omega_D/\varepsilon_F \ll 1$), we use ${\cal N}=100$ for all $\lambda$.

In the non-adiabatic regime at weak coupling, convergence is slower than in the adiabatic case. For $\omega_D/\varepsilon_F = 10, 100$ and $\lambda = 0.05$, it takes ${\cal N} = 250$ to reach convergence of $T_c$, as shown in Fig.~\ref{fig:S2} (left panel). For $\lambda=1$, ${\cal N} =10$ is sufficient for the convergence of $T_c$, as seen in Fig.~\ref{fig:S2} (right panel). To compute $T_c$ for $\omega_D/\varepsilon_F\geq 1$, we use ${\cal N}=300$ for $\lambda<1$ and ${\cal N}=100$ for $\lambda\geq1$.

From Fig.~\ref{fig:S2}, we note that the critical temperatures for large adiabatic ratios ($\omega_D/\varepsilon_F = 10, 100$) coincide, indicating that the $T_c$ curves will merge for $\omega_D/\varepsilon_F \gg 1$ if one measures $T_c$ in units of $\varepsilon_F$.

\section{IV. Critical temperature in the strong-coupling limit in the absence of the bosonic self-energy}

Consider the system of the Migdal-Eliashberg theory equations (4,5,9,10) in the main text but without a self-consistent account of the bosonic self-energy. Instead, we use a fixed bosonic propagator for an Einstein boson of mass $\epsilon$:
\begin{equation}
{\cal D}(i\Omega_m) = \frac{1}{\Omega_m^2 +\epsilon^2}.
\label{D_E}
\end{equation}
A similar effective bosonic propagator is often considered in the conventional Migdal-Eliashberg theory and corresponds to the delta function-like spectral density of bosons~\cite{Carbotte1990}. Although there is no explicit notion of the adiabatic ratio in the bosonic propagator, the adiabatic approximation ($\epsilon/\varepsilon_F \ll 1$) is preliminarily assumed since we have neglected the self-energy of the boson. To state the slow-boson condition for our model, we use $\epsilon$ rather than $\omega_D$.

Retracing the steps of the main text with the new bosonic propagator \eqref{D_E}, we obtain~\cite{Bennemann}:
\begin{align} 
z(i\omega_n) \Delta(i\omega_n) &= \pi \lambda T_c \sum_{n'=0}^{+\infty} \frac{\Delta(i\omega_{n'})}{\omega_{n'}}  \bigg[(1\!-\!\delta_{nn'}) L(i\omega_n\!-\!i\omega_{n'}) \!+\! L(i\omega_n\!+\!i\omega_{n'})  \bigg], \label{T_c_eq_E} \\
z(i\omega_n) &= 1 + 2\pi \lambda  \,\frac{T_c}{\omega_n} \sum_{m=1}^n L(i\Omega_m), \label{z_E}
\end{align}
where $L(i\Omega_m) = \epsilon^2 {\cal D}(i\Omega_m)$ and $\lambda = \nu g^2/\epsilon^2$ is the electron-boson coupling constant. Performing the summation over bosonic frequencies in the renormalization function \eqref{z_E}, we have
\begin{equation}
    z(i\omega_n) =  1 \!+\! \lambda \, \frac{\epsilon}{\omega_n} {\rm Im}\left[\psi\!\left(1\!+\!\frac{i\epsilon}{2\pi T_c}\right) - \psi\!\left(1 \!+\! n \!+\!\frac{i\epsilon}{2\pi T_c}\right) \right], \label{z_E_sum}
\end{equation}
where $\psi(z) = -\gamma - \underset{n=0}{\overset{+\infty}{\sum}}[1/(n+z) - 1/(n+1)]$ is the digamma function and $\gamma$ is the Euler-Mascheroni constant. For $\epsilon \ll 2\pi T_c$ \cite{Carbotte1990}, we expand Eqs.~(\ref{T_c_eq_E},\ref{z_E_sum}) to second order in $\epsilon/T_c$. As a result, the pairing problem simplifies to 
\begin{align} 
z(i\omega_n)\Delta(i\omega_n) &= \pi \lambda \epsilon^2 T_c \sum_{n'=0}^{+\infty} \frac{\Delta(i\omega_{n'})}{\omega_{n'}}  \bigg[\frac{1-\delta_{nn'}}{(\omega_n-\omega_{n'})^2} \!+\! \frac{1}{(\omega_n+\omega_{n'})^2}\bigg], \label{T_c_eq_E_sc} \\
z(i\omega_n) &= 1 +\! \lambda \, \frac{\epsilon^2}{2\pi T_c \omega_n} H_{n,2}. \label{z_E_sc}
\end{align}
In the above we used that $\psi^{(1)}(1)-\psi^{(1)}(1+n)=H_{n,2}$, where $\psi^{(1)}(z)=d\psi(z)/dz$ and $H_{n,2}$ is the harmonic number of the second order. 

Let us solve the characteristic equation (13) in the main text for the pairing problem (\ref{T_c_eq_E_sc}-\ref{z_E_sc}) with the first two Matsubara frequencies ($n,n'=0,1$). To do so, it is convenient to measure $T_c$ in units of $\epsilon\sqrt{\lambda}$: $\overline{T}\equiv T_c/(\epsilon\sqrt{\lambda})$. Then, we find the critical temperature from 
\begin{equation}
    \begin{vmatrix}
        \dfrac{1}{4(\pi \overline{T})^2} -1 & \dfrac{5}{48(\pi \overline{T})^2} \\ \dfrac{5}{\left(1+\frac{1}{6(\pi\overline{T})^2}\right)16(\pi \overline{T})^2} & \dfrac{1}{\left(1+\frac{1}{6(\pi\overline{T})^2}\right)108(\pi \overline{T})^2}-1
    \end{vmatrix} = 0. \label{det2x2}
\end{equation}
Eq.~\eqref{det2x2} has a single real positive root, $\overline{T}= \frac{1}{12\pi}\sqrt{\frac{1}{3}(20+\sqrt{13819})}\approx0.18$~\cite{Kresin1984Superconducting}, which reproduces the Allen-Dynes result $T_c \approx 0.18\, \epsilon \sqrt{\lambda}$~\cite{Allen1975Transition}. Inclusion of the third Matsubara frequency in the characteristic equation gives $T_c \approx 0.182\, \epsilon \sqrt{\lambda}$, while adding the fourth Matsubara frequency specifies the third digit after the decimal point:
\begin{equation}
    T_c \approx 0.183 \, \epsilon \sqrt{\lambda}. \label{T_c_E_AD}
\end{equation}
In the limit that $\lambda \to \infty$, Eq.~\eqref{T_c_E_AD} is consistent with the small boson mass approximation that we made.

\begin{figure}[t!]
\center
\includegraphics[width=0.37\linewidth]{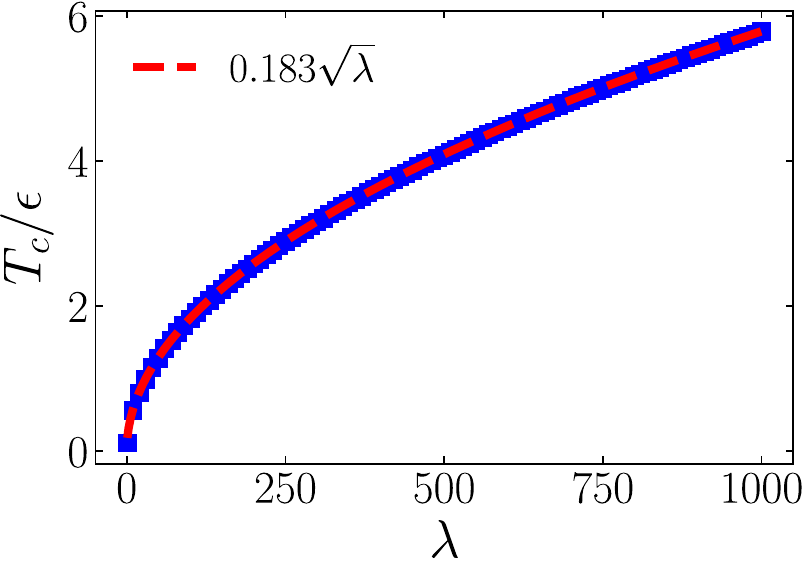}
\caption{\small \label{fig:S3} {\bf Numerical solution for $T_c$ in the absence of the self-energy of the boson for $\lambda\geq 1$}. The blue squares show the numerical result from Eq.~\eqref{T_c_eq_E}. The dashed red curve shows  $T_c$ from the strong-coupling formula \eqref{T_c_E_AD}.}
\end{figure}

In Fig.~{\ref{fig:S3}}, we show $T_c$ as calculated numerically from the exact pairing problem (\ref{T_c_eq_E}-\ref{z_E}) with $100$ Matsubara frequencies and a step size of $10^{-2}\,\epsilon$ for $\lambda \geq 1$. The numerical result agrees with Eq.~\eqref{T_c_E_AD}.
 
In the absence of the self-energy of the boson, there is no mechanism to suppress the growth of $T_c$, which diverges in the strong-coupling limit.

\bibliography{refs}